\documentclass[nolineno,a4paper,UKenglish,cleveref, autoref, thm-restate,]{lipics-v2021}

\hideLIPIcs
\hypersetup{pdfsubject={}}

\usepackage{cite}
\usepackage{amsmath,amssymb,amsfonts}
\usepackage{textcomp}
\usepackage{xcolor}	
\usepackage{soul}
\usepackage{spverbatim}
\usepackage{verbatimbox}
\usepackage{graphicx}
\usepackage[shortlabels]{enumitem}
\usepackage{bytefield}
\usepackage{algorithm}
\usepackage{float} 
\usepackage{hyperref}
\usepackage{amsmath}
\usepackage{pgfplots}
\pgfplotsset{compat=1.18}
\usepgfplotslibrary{groupplots}
\usepgfplotslibrary{fillbetween}
\usepackage[table]{xcolor}
\usepackage[noend]{algpseudocode}
\usepackage{subcaption}
\usepackage{siunitx}
\usepackage[table]{xcolor}
\definecolor{lightgreen}{RGB}{16,181,77}
\usepackage{tikz}

\newcommand{\smallcircled}[1]{\tikz[baseline=(char.base)]{
    \node[shape=circle,draw,inner sep=0.1pt, minimum size=7pt, font=\scriptsize] (char) {#1};}}
    
\newcommand{\circled}[1]{\tikz[baseline=(char.base)]{
    \node[shape=circle,draw,inner sep=0.5pt, font=\footnotesize] (char) {#1};}}

\newcommand{\blackdot}[1]{%
  \tikz[baseline=(char.base)]{%
    \node[shape=circle,draw=black,fill=black,text=black,inner sep=0pt] (char) {#1};%
  }%
}

\newenvironment{tinyalgorithmic}{\begin{algorithmic}[1]\tiny}{\end{algorithmic}}

\newcommand{\instruction}[1]{{\texttt{#1}}}
\newcommand{\event}[1]{{\textit{#1}}}
\newcommand{\type}[1]{{\textit{#1}}}

\newcommand{\condition}[1]{{\textit{#1}}}
\newcommand{\variable}[1]{{$#1$}}

\newboolean{showcomments}
\setboolean{showcomments}{true}
\ifthenelse{\boolean{showcomments}}
{ \newcommand{\mynote}[3]{
    \fbox{\bfseries\sffamily\scriptsize#1}
    {$\blacktriangleright$\textsf{{\color{#3}{#2}}}$\blacktriangleleft$}}}
{ \newcommand{\mynote}[3]{}}

\newcommand{\new}[1]{\textcolor{purple}{#1}}

\newcommand{\subheading}[1]{\par\vspace{0.4em}\noindent\textbf{\normalsize #1.}}

\makeatletter
\newcommand{\heading}[1]{%
  \par\vspace{0.4em}\noindent\large\bfseries #1\par\vspace{0.3em}\normalfont\normalsize
  \@afterindentfalse\@afterheading
}
\makeatother

\newcommand{\LONG}[1]{}
\newcommand{\VTWO}[1]{}

\def\BibTeX{{\rm B\kern-.05em{\sc i\kern-.025em b}\kern-.08em
    T\kern-.1667em\lower.7ex\hbox{E}\kern-.125emX}}

\title{Dora and Dave: Learning to Synthesize Verified Broadcast Algorithms} 

\author{Diogo Vaz}{INESC-ID, Instituto Superior Técnico, Lisboa -- Portugal}{}{}{}

\author{David R. Matos}{INESC-ID, Instituto Superior Técnico, Lisboa -- Portugal}{}{}{}

\author{Miguel L. Pardal}{INESC-ID, Instituto Superior Técnico, Lisboa -- Portugal}{}{}{}

\author{Miguel Correia}{INESC-ID, Instituto Superior Técnico, Lisboa -- Portugal}{}{}{}

\authorrunning{D. Vaz et al.} 

\Copyright{Diogo Vaz, David R. Matos, Miguel L. Pardal and Miguel Correia} 

\ccsdesc{Computing methodologies~Distributed algorithms} 

\keywords{Distributed Algorithms, 
Broadcast, Program Synthesis, Monte Carlo Tree Search, Program Verification, 
Model Checking, 
SPIN} 

\acknowledgements{This work was supported by Funda\c{c}\~{a}o para a Ci\^{e}ncia e a Tecnologia with references UIDB/50021/2020 (INESC-ID) and 2022.10788.BD and by Rede Nacional de Computação Avançada with references 2025.09017.CPCA.A1 and 2025.13378.CPCA.A1.}

\begin{document}

\renewcommand{\arraystretch}{0.9}
\setlength{\tabcolsep}{4pt}
\nolinenumbers
\hypersetup{pdfsubject={}}

\maketitle

\begin{abstract}
Distributed computing relies on fault-tolerant algorithms to ensure reliable operation despite failures. However, designing such algorithms is complex and time-consuming, which has motivated recent approaches to automate this process. These efforts have remained limited in scale and scope, as the enormous search space, with tens of millions of possible candidate solutions, inherent in distributed algorithm design, poses a major scalability obstacle. To address this challenge, we introduce a verifier-guided test-time search approach over a grammar-constrained program space, based on a per-run online learning process, rather than relying on a pretrained neural model. Our approach is supported by two agents: \textsc{Dora}, an efficient self-learning agent for the automatic synthesis of distributed algorithms, and \textsc{Dave}, a formal verifier agent that automatically verifies the algorithms. \textsc{Dora} explores design spaces by generating algorithms in parallel, combining Monte Carlo Tree Search with a flexible verifiable reward mechanism, enabling effective exploration of design spaces. We apply our approach to different broadcast problems such as Reliable Broadcast, Uniform Reliable Broadcast, and Terminating Reliable Broadcast. The experimental results show that \textsc{Dora} and \textsc{Dave} synthesize algorithms that rival the manually designed algorithms within one day of computation. Compared with the previous single machine-learning-based approach, \textsc{Dora} identifies up to $18\times$ more correct algorithms while exploring only $3\times$ more states, demonstrating substantially improved sample efficiency.
\newline
\newline

\end{abstract}


\newpage
\section{Introduction}
\label{sec:intro}
\vspace{-0.2cm}

Fault-tolerant distributed algorithms~\cite{cachin_introduction_2011,lynch_distributed_1996,raynal_fault_2018} enable multiple machines to operate 
as a coordinated distributed system~\cite{coulouris_distributed_2005}, 
by ensuring correct behavior even under failures.  

Despite remarkable progress~\cite{bonomi_practical_2021,bracha_asynchronous_1984,castro_practical_1999,correia_consensus_2006,raynal_consensus_2002}, the design of distributed algorithms is complex and expensive~\cite{hadzilacos_modular_1994,mickens_saddest_2014}. Even small changes, e.g., modifying failure assumptions or synchrony model, may require redesigning the protocol. For example, when solving \emph{Reliable Broadcast} in an asynchronous system, changing failures from crash to arbitrary not only roughly doubles the messages exchanged~\cite{cachin_introduction_2011,bracha_asynchronous_1984}, but also
strengthens the resilience requirement from $N>F$ to $N>3F$. In this work, we focus on \emph{Broadcast}~\cite{cachin_introduction_2011,hadzilacos_modular_1994} protocols, known for their practical application~\cite{birman_reliable_1985,verissimo_reliable_1990,cristian_fault_1990} and for being important building blocks for more complex algorithms, such as \emph{Consensus}~\cite{hadzilacos_modular_1994,bracha_asynchronous_1984,cachin_introduction_2011}. Therefore, improving the efficiency of broadcast primitives can benefit entire classes of distributed protocols and systems. Nevertheless, viewing broadcast protocols as primitives does not diminish their inherent complexity; prior work proved \emph{Terminating Reliable Broadcast} and \emph{Consensus} to be equivalent in synchronous systems~\cite{hadzilacos_modular_1994}.

To assist in the design of distributed algorithms, prior research has explored automatic synthesis using brute force~\cite{bar_automatic_2003}, model-checking-based genetic programming~\cite{katz_model_2008,katz_genetic_2008,katz_synthesizing_2011}, enumerative~\cite{zielinski_automatic_2007}, Satisfiability Modulo Theories (SMT)-based bounded model checking~\cite{delporte_towards_2019}, and Counterexample-Guided Inductive Synthesis (CEGIS)~\cite{lazic_synthesis_2018} techniques.
Nevertheless, automatic synthesis remains extremely challenging due to the enormous search space inherent to distributed protocol design, which can easily reach tens of millions of candidate behaviors~\cite{bar_automatic_2003,zielinski_automatic_2007,delporte_towards_2019}. As a result, existing techniques are typically restricted to small, fixed-size instances~\cite{bar_automatic_2003,delporte_towards_2019}, partial algorithm templates~\cite{lazic_synthesis_2018}, or optimistic assumptions~\cite{zielinski_automatic_2007}.
For this reason, researchers have more recently turned to machine learning, particularly reinforcement learning, to guide distributed algorithm generation~\cite{vaz_automatic_2023,vaz_synthesis_2023,hui_learning_2026}. Yet prior machine-learning-based approaches have each targeted a narrow set of problems, namely, reinforcement learning has been applied only to asynchronous \emph{Reliable Broadcast}~\cite{vaz_synthesis_2023,vaz_automatic_2023}. \LONG{, \new{while MCTS combined with model checking has targeted only synchronous Consensus and atomic commit under crash failures~\cite{hui_learning_2026}.}}



To address these limitations, we propose \textsc{Dora}, a \emph{\underline{D}istributed alg\underline{o}rithms generato\underline{r} \underline{a}gent}, built around \emph{a verifier-guided test-time search technique}. All used statistics are built online, within a single run, where candidate algorithms are generated and verified for the problem instance at hand, rather than relying on a model pretrained offline on a large corpus. We formulate the synthesis of algorithms as a structured search problem, modeled as a single-player \underline{B}roadcast \underline{G}ame (BGame) in which \textsc{Dora} searches for correct and efficient solutions on a constrained state space of partial and complete candidate algorithms/programs constructed from a defined language/grammar of structures and actions. A second agent, \textsc{Dave}, a \emph{\underline{D}istributed \underline{a}lgorithms \underline{v}erifier ag\underline{e}nt}, verifies the correctness of the solutions. We employ \emph{model checking}~\cite{clarke_model_2018,baier_principles_2008} to verify the algorithms (see Section~\ref{section:verification_process}) and show the proofs of correctness for the optimal algorithms (see Appendix~\ref{appendix:correctness_proofs_2}.)


Our approach introduces three techniques compared to previous work. First, it follows a learning technique based on \emph{Monte Carlo Tree Search} (MCTS)~\cite{browne_survey_2012,coulom_efficient_2006}, allowing the exploration of large state spaces. Second, it employs a parallel generation architecture that increases coverage of the state space. Third, it introduces a flexible and verifiable reward mechanism that improves search efficiency by guiding exploration toward promising space regions.

Our results show that \textsc{Dora} and \textsc{Dave} generate multiple crash- and omission-tolerant broadcast algorithms that rival handcrafted solutions under the evaluated metrics, demonstrating it can synthesize correct and efficient distributed algorithms. 
Under the same time budget and in the same state space, our approach discovers up to $18\times$ more correct algorithms while exploring only $3\times$ more states than the prior machine-learning system~\cite{vaz_automatic_2023,vaz_synthesis_2023}, 
with the best 3 algorithms found 
improving the mean reward by up to \new{$52\%$}. 


The main contributions of this paper are:
\circled{1} A new verifier-guided test-time approach for automatically synthesizing distributed algorithms, combining an MCTS-based learning technique, a parallel generation architecture, and a flexible verifiable reward mechanism; 
\circled{2} An application of this approach to synthesizing multiple broadcast algorithms, producing crash- and omission-tolerant solutions, competitive with human-designed algorithms.

\vspace{-0.3cm}
\section{BGame: Generating Broadcast Algorithms}
\label{section:trb_game}
\vspace{-0.2cm}

\emph{BGame} (\underline{B}roadcast \underline{G}ame) is a single-player game in which \textsc{Dora} incrementally selects actions to generate a candidate broadcast algorithm, queries \textsc{Dave} to determine whether the candidate is correct, and receives a reward based on correctness and efficiency. The goal is to discover the algorithm that maximizes this reward. Next, we introduce the key concepts.



\heading{2.1 Goal}
\label{section:goal}
The goal is to find a correct and efficient solution to the broadcast problems. Next, we explain the \emph{properties}, \emph{efficiency}, and the \emph{system models} considered in this work.

\subheading{2.1.1 Properties}
\label{section:properties}
\begin{table}\scriptsize
  \caption{Properties supported by our system.}
  \label{tab:properties}
  \hspace*{-0.7cm}
  \centering
  \begin{tabular}{l | l}
  \hline
    \textbf{Property} & \textbf{Definition} \\
    \hline
    \textit{Validity} &
      if a correct process \event{Broadcast}s a value $v$, then (1) it will eventually deliver some value  \\
      & and (2) the value cannot differ from $v$. \\
    \hline
    \textit{Agreement} &
      if two correct processes deliver, they do not deliver distinct values. \\
    \hline
    \textit{Uniform Agreement} &
      if two processes deliver, either correct or not, they do not deliver different messages. \\
    \hline
    \textit{No-duplication} &
      if a correct process delivers a value $v$, it does not deliver $v$ more than once. \\
    \hline
    \textit{Integrity1} &
      if a correct process delivers a value $v$, then $v$ was \event{Broadcast} by some process. \\
    \hline
    \textit{Integrity2} &
      if a correct process delivers a value $v$, then $v = SF$ \emph{(Sender Faulty)} or $v$ was \event{Broadcast} \\
      & by some process. \\
    \hline
    \textit{Totality} &
      if one correct process delivers a value, then all correct processes eventually deliver some value. \\
    \LONG{\textit{Best-Effort Totality} &
      if the broadcaster is correct and one correct process delivers a value, then all correct processes eventually deliver some value. \\}
    \hline
    \textit{Uniform Totality} &
      if one process, either correct or not, delivers a value, then all correct processes eventually \\
      & deliver some value. \\
    \hline
    \emph{Termination} &
      every correct process eventually delivers some value. \\
      \hline
  \end{tabular}
\end{table}
Distributed algorithms are often specified in terms of sets of properties \cite{cachin_introduction_2011,raynal_fault_2018}. In this paper we consider properties often used to define Broadcast problems, shown in Table~\ref{tab:properties}. For other problems, e.g., Consensus, others would be added.

\subheading{2.1.2 Efficiency}
\label{section:efficiency}
To analyze the efficiency of an algorithm, we adopt a model inspired by previous work \cite{cachin_introduction_2011, mostefaoui_signature_2015,raynal_consensus_2002,vaz_synthesis_2023}, where the efficiency is assessed in terms of three metrics:  
\circled{1} the \emph{number of messages sent} by the algorithm; 
\circled{2} the \emph{number of messages needed to reach a decision}
; and 
\circled{3} the \emph{number of message types}, as its use typically implies additional messages and \emph{communication steps}. 
All of these metrics indirectly express the cost of computational power, storage, and network resources needed to execute the algorithm.

\subheading{2.1.3 System Model}
\label{section:system_model}
We assume that the algorithms generated will run on a \emph{message-passing} system 
composed of a \emph{static group} of $\emph{N} \in \mathbb{N}$ computation nodes designated as \emph{processes} 
on a \emph{fully connected point-to-point} network. 
We assume that the links are \emph{reliable}, \emph{authenticated}, and \emph{provide message integrity}, i.e., there are no corrupted, lost, impersonated, or duplicate messages. Messages may arrive out of order or not arrive if the sender or receiver fail. We consider two timing models: \emph{synchronous}, where known timing bounds exist for computation and communication, and \emph{asynchronous}, where no such bounds are assumed. As future work, we plan to extend the support to partially synchronous systems, in which time bounds exist but are unknown. 

We assume only processes can fail, with $\emph{F} \in \mathbb{N}$ defining the maximum number of faulty nodes. In this synthesis process, both $N$ and $F$ are configuration parameters fixed at synthesis and verification time. During verification, Dave exercises every actual fault count from $0$ up to $F$ (Section~\ref{section:verification_process}). We consider the \emph{crash-failure} model, where processes may stop operating and never recover, and the \emph{general omission-failure} model, where processes may omit to send and/or receive messages but cannot fabricate, impersonate or corrupt values. Timing failures~\cite{hadzilacos_modular_1994}, where messages may arrive outside of the timing bound, are out of scope. 


\heading{2.2 Board}
\label{section:board}
The board of the game, where the selected pieces are added, is 
the template of the algorithm that we want to generate (see Algorithm~\ref{algorithm:algorithm_structure}). 
The two main components of the board/algorithm are \emph{events} and \emph{variables}.

\begin{figure}[th]
\begin{minipage}[t]{0.52\textwidth}
\vspace{-0.7cm}
\begin{algorithm}[H]
\caption{Template for generating the algorithms.}
\vspace{-0.1cm}
\label{algorithm:algorithm_structure}
\begin{tinyalgorithmic}
\State \textbf{// Inner state:}
\State $ID \in {0,...,N-1}$       \Comment{// process identifier}
\State $msg\_received := \varnothing$      \Comment{// set of received messages, initially empty}
\State $msg\_sent := \varnothing$           \Comment{// set of messages sent, initially empty}
\State $delivered\_value := \perp$    \Comment{// initially undefined}
\State $default\_value := SF$
\State 
\State \textbf{// Algorithm:}
\State \event{when Broadcast($v$)}:
\State \hskip1.0em
\color{gray}
/*\textsc{Dora} adds actions here*/
\color{black}
\State \event{end;}
\State \event{when Receive($v$,$t$,$f$) do}: 
\State \hskip1.0em
\color{gray}
/*\textsc{Dora} adds actions here*/
\color{black}
\State \event{end;}
\State \event{when $\Delta-$Timeout}:
\State \hskip1.0em 
\color{gray}
/*\textsc{Dora} adds actions here*/
\color{black}
\State \event{end;}
\vspace{-0.1cm}
\end{tinyalgorithmic}
\end{algorithm}
\end{minipage}
\hfill
\begin{minipage}[t]{0.44\textwidth}
\vspace{-0.8cm}
\begin{table}[H]
\tiny
\caption{Instructions and conditions considered for the generation process.} 
\vspace{-0.2cm}
\centering 
\begin{tabular}{l | r }
\hline 
 \textbf{Instructions} & \textbf{Cost} \\ [0.5ex]
\hline
 \instruction{Send to N Processes}(\variable{value},\variable{type},\variable{sender}) & $3$ \\
 \instruction{Send to N-1 Neighbors}(\variable{value},\variable{type},\variable{sender}) & $2$ \\
 \instruction{Deliver}(\variable{value}) & $1$ \\
 \instruction{Stop}() & $0$ \\
 \hline
 \textbf{Conditions} & \textbf{Cost} \\
 \hline
\condition{ReceivedEqualMsg}(\variable{value},\type{type},\variable{threshold}) & $1$ \\
\condition{IsTrue()} & $0$ \\
\hline
\end{tabular}
\label{table:instructions_conditions}
\end{table}
\end{minipage}
\end{figure}



\subheading{2.2.1 Events}
\label{section:events}
Events are routines containing a set of actions executed by specific triggers. This \emph{event-based} structure has been widely used in the design of distributed algorithms~\cite{bonomi_practical_2021,cachin_introduction_2011,moniz_randomization_2011,guerraoui_dynamic_2021}, including automatic synthesis approaches~\cite{vaz_synthesis_2023}. 
The template has three events: \event{Broadcast}, \event{Receive}, and $\Delta$\event{-Timeout}. \event{Broadcast} (line $9$ of Algorithm \ref{algorithm:algorithm_structure}) is the event that initiates the protocol, triggered by a single process, the \emph{broadcaster}, wanting to share a value. The \event{Receive} event (line $12$) is triggered every time a process receives a message. The \event{$\Delta$-Timeout} (line $15$), only available for synchronous algorithms, is a timing event that each process arms at the common start time $T_0$ known by all (e.g., through synchronized clocks, a standard assumption in synchronous systems~\cite{cachin_introduction_2011}). The bound $\Delta$ is defined to exceed the worst-case end-to-end execution of the algorithm, accounting for combined communication delays, local computation, and possible deviations. We also assume the execution of each event is atomic: while an event is being executed, no other event can be processed.

\subheading{2.2.2 Variables}
\label{section:variables}
Variables represent the process's internal state, based on those commonly used by human-generated algorithms~\cite{schiper_early_1997,moniz_randomization_2011,cachin_introduction_2011}. Variables have one of two scopes: \emph{algorithm scope} (accessed throughout the algorithm) or \emph{event scope} (accessed only within its event). In \emph{algorithm scope} (lines $2$-$6$), we define five variables: $ID$, $msg\_received$, $msg\_sent$,  $delivered\_value$ and $default\_value$. 
$ID$ identifies the process. $msg\_received$ stores received messages, needed to check threshold conditions (Section~\ref{section:conditions}). $msg\_sent$ stores sent messages, preventing duplicate sends (optional, since the recipient can discard duplicates). $delivered\_value$ stores the decided value. $default\_value$ stores the default value $SF$ (Sender Faulty): a special message delivered when the initial value is unavailable. In \emph{event scope}, the \event{Broadcast} event's local variable $v$ contains the value to propose. In the \event{Receive} event, every received message carries $3$ values, represented as local variables $v$ (the value), $t$, and $f$. $t$ is the message type, differentiating communication steps/phases; \textsc{Dora} generates types \type{mtype0}, \type{mtype1}, \type{mtype2}, etc. automatically. $f$ is the message's sender ID.


\heading{2.3 Actions}
\textsc{Dora} solves the game by selecting pieces to add to the board, i.e., actions to add to the algorithm. Table \ref{table:instructions_conditions} presents the information used to build the actions. Each action has two components: \emph{instruction}, the part of the action that is executed depending on specific \emph{arguments}, and \emph{conditions}, statements that, depending on specific \emph{arguments}, must be true for the instruction to be executed. Each action is defined as:

\vspace{-0.1cm}
\begin{center}\scriptsize
if \textcolor{violet}{\condition{condition}(\variable{arguments})} do \textcolor{brown}{\instruction{instruction}(\variable{arguments})}, e.g. if \textcolor{violet}{\condition{isTrue()}} do \textcolor{brown}{\instruction{Send to N Processes(\variable{v}, \variable{mtype0},\variable{ID}})}
\end{center}
\vspace{-0.1cm}


\noindent
In the example, the condition component is the left-hand part (in \textcolor{violet}{violet} starting with ``if'') and the instruction is the right-hand part (in \textcolor{brown}{brown} after the word  ``do''). This action means that the process will send a message of $mtype0$ with the value \variable{v} to $N$ processes, i.e., the \condition{isTrue} condition is an \emph{always} true state, meaning that the instruction will always be executed. Next, we present the instructions and conditions 
that we use.

\subheading{2.3.1 Instructions}
\label{section:instructions}
Instructions are statements that define the behavior of the process. Each instruction can receive an arbitrary number of arguments. Table \ref{table:instructions_conditions} presents all the instructions defined
. These instructions include the \emph{sending} and \emph{delivering} of messages. The \instruction{Send to N-1 Neighbors} sends a message to all neighbors, excluding the sender. 
The \instruction{Send to N Processes} instruction sends a message to all the system processes, including the sender itself. 
All \instruction{Send} instructions automatically avoid duplicate messages: none sends the same message, with the same type, to the same process, through the $msg\_sent$ variable. The \instruction{Deliver} instruction allows the process to deliver a message by storing a value on the $delivered\_value$ variable and terminates the execution of the algorithm. Finally, the \instruction{Stop} instruction means the end of the execution of the event.


\subheading{2.3.2 Conditions}
\label{section:conditions}
Conditions evaluate when an instruction is to be executed. We define two types of conditions: \circled{1} \condition{IsTrue}, a neutral condition representing an always true behavior; and \circled{2} \condition{ReceivedEqualMsg}, for waiting for a specific \variable{threshold} 
of a specific \variable{type}. 
A \variable{threshold} variable specifies the minimum number of messages of a given \variable{type} that must be received before the instruction can be executed. We automatically explore meaningful thresholds by adopting a technique similar to 
Lazić et al.~\cite{lazic_synthesis_2018}, defining the \variable{threshold} as: 

\vspace{-0.2cm}
{\scriptsize
$$
threshold(N,F) = Q_a/d \cdot N + Q_b/d \cdot F + Q_c 
\quad \text{with}
$$
$$
0 \leq Q_a \leq d\cdot1, \quad
-(r + 1)\cdot d \leq Q_b \leq (r + 1) \cdot d, \quad -(2 + 2\cdot r)\cdot d \leq Q_c \leq (2 + 2\cdot r)\cdot d
$$
}
\vspace{-0.6cm}

\noindent where $N$ is the number of processes, $F$ is the maximum of faults, $r$ the system's resilience and $d$ the maximum coefficient denominator. \LONG{For example, assuming a system's resilience condition $r=1$, where $N > F$, and adopting $d=3$ as a practical upper bound, we derive bounds for the search space:}
\LONG{
{\scriptsize
$$
0 \leq Q_a \leq 3, \quad -6 \leq Q_b \leq 6, \quad -12 \leq Q_c \leq 12
$$
}
}
Then, we apply a set of filters (\variable{threshold} $ \geq 1$ and \variable{threshold} $ \leq N$), to exclude non-viable or redundant thresholds (e.g., negative or greater than $N$), ensuring \textsc{Dora} only explores a curated catalog of meaningful threshold conditions. \LONG{In this example, $22$ thresholds would be created (i.e., $1$, $F+1$, $N-1$, $N-F$, $N - F/3$ to name a few).}

In the end, to create the set of possible actions
, \textsc{Dora} combines all instructions with all conditions except for the \instruction{Stop} instruction: it was only combined with the \condition{IsTrue} condition. By design, \textsc{Dora} is restricted to a fixed catalog of actions: candidate algorithms can only be built by combining these instructions and conditions. This keeps the space well-defined and avoids the unpredictability of unconstrained generation, while still allowing \textsc{Dave}'s verification to give reliable feedback on every candidate. For more expressiveness, additional instructions and conditions can be added.




\heading{2.4 Playing the game}
\label{section:playing_the_game}

\textsc{Dora} generates an algorithm by incrementally adding actions to the template (Algorithm~\ref{algorithm:algorithm_structure}), starting from an empty algorithm and developing one event at a time: \event{Broadcast}, then \event{Receive}, then $\Delta$\event{-Timeout} (if applicable). At each step, \textsc{Dora} collects all possible actions and applies the \emph{heuristics} (Section~\ref{section:heuristics}) to discard those unsuitable for the current state, then selects an action from the remaining subset and adds it to the current event. The agent advances to the next event after a \instruction{Stop} action is chosen, and the algorithm completes once all events contain \instruction{Stop}. A detailed explanation, with pseudo-code, can be found on Appendix~\ref{appendix:building_an_algorithm}. 

\textsc{Dora} generates a single algorithm per game, but producing a \emph{correct and efficient} one requires \emph{learning} across multiple games. Next, we explain \textsc{Dora}'s learning process.


\section{Learning to Synthesize Efficient Algorithms}
\vspace{-0.2cm}

\textsc{Dora} computes its decisions using the \emph{Monte Carlo Tree Search} (MCTS) algorithm~\cite{browne_survey_2012,coulom_efficient_2006}. MCTS is a simulation-based planning method that incrementally constructs a search tree of states and actions by repeatedly simulating possible future trajectories using a model of the environment. Each simulation traverses the tree by balancing exploration and exploitation, expands new nodes when unexplored states are encountered, and propagates the obtained reward back to update the estimated return of the visited nodes. 
With time, the accumulated simulations will allow \textsc{Dora} to refine and progressively improve the quality of action selection.

In the following sections, we present \textsc{Dora}, its architecture, and organization.

\begin{figure}[th]
    \centering
    \includegraphics[width=0.65\columnwidth]{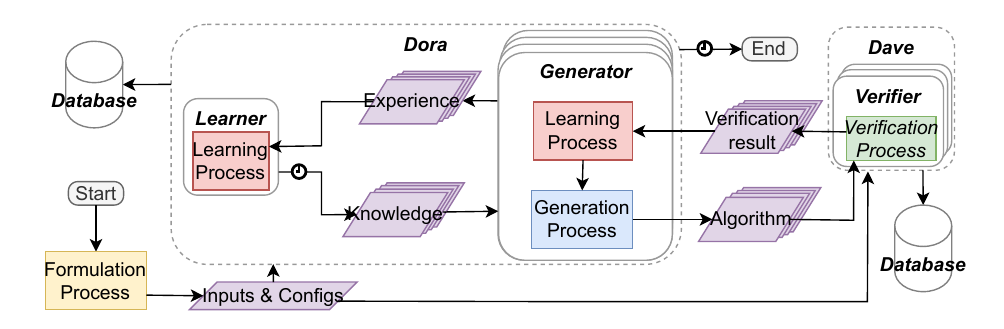}
    \caption{Process/dataflow of the system. Notice that there can be several generators and verifiers.}
    \label{fig:solution_architecture}
\end{figure}

\heading{3.1 Multi-algorithm generation and the agents} 
\label{section:solution}
We aim to synthesize correct and efficient 
broadcast algorithms, i.e., algorithms that satisfy specific properties 
and minimize the efficiency metrics (Section \ref{section:efficiency}). To achieve this, we present two agents: \emph{Dora} and \emph{Dave}. Figure \ref{fig:solution_architecture} presents the architecture of our solution, organized on a multi-agent architecture: we assume the existence of a single \textsc{Dora} agent and multiple \textsc{Dave} agents (due to computational constraints, we had to run multiple \textsc{Dave}s on independent machines). Next, we present each agent.

\subheading{3.1.1 \textsc{Dora}}
This agent contains two main components, the \emph{Generator}, 
 and the \emph{Learner}, organized in a multi-component setting: one \emph{Learner} and multiple \emph{Generators}. This architecture is based on Root Parallelization MCTS~\cite{chaslot_parallel_2008} and previous works that provide multi-agent architectures to speed up the learning process
~\cite{espeholt_impala_2018,horgan_distributed_2018}, by implementing independent generation and learning processes, where multiple actors generate \emph{experience} consumed by a single learner. Then, periodically, the learning agent shares the knowledge with the generation agents. In this work, 
we propose multiple \emph{Generators} that generate multiple algorithms in parallel, and a unique \emph{Learner} learning from all the experience generated. Then, periodically, the Learner component shares its knowledge with each Generator. Note that our goal is not to present a new parallel algorithm
, but to implement one that allows us to improve our solution when compared to using a single-component architecture. Moreover, we stress that learning is online and local to each synthesis run -- a full execution of \textsc{Dora} and \textsc{Dave} devoted to a single problem instance -- and that \textsc{Dora} carries no model or parameters across problem instances and undergoes no offline training. Therefore, ``learning'' refers to the MCTS statistics (cf. Section 3.2.3) that \textsc{Dora} refines as the search progresses.

\subheading{3.1.2 \textsc{Dave}}
\textsc{Dave} is not a monolithic component but a set of \emph{Verifiers} that verify the correctness of the algorithms generated (\emph{Verification Process}). Algorithms are thus generated and verified in parallel. \textsc{Dora} and \textsc{Dave} share a database server where they store information on the algorithms generated and their verification status.





\heading{3.2 Generation process}
\label{section:generation_process}
\begin{figure}[th]
    \hspace*{1.4cm}
    \includegraphics[width=0.7\columnwidth]{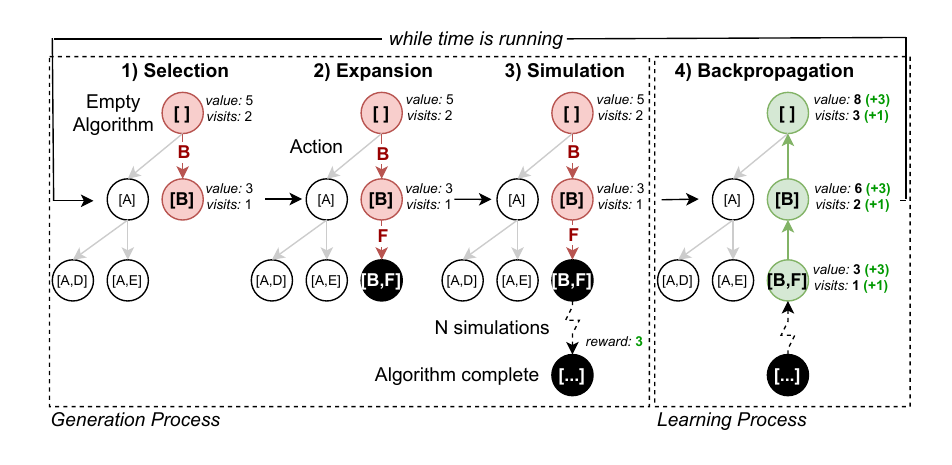}
    \caption{Monte Carlo Tree Search algorithm. From the root state $[\space]$, actions $B$ and $F$ are selected using the \emph{Upper Confidence Bound for Trees} (UCT) (see Section 3.2.1) policy, leading to the $leaf$ state [$B$,$F$]. Since state [$B$,$F$] is new, $N$ simulations are run to find the value of state [$B$,$F$]. The result from the simulations is propagated from the leaf state [$B$,$F$] to the root state $[\space]$.}
    \label{fig:monte_carlo_process}
\end{figure}

Our solution starts with the manual \emph{Formulation Process}: the definition of the \emph{inputs} and \emph{configurations}. For \textsc{Dora}, we define the total execution time, the rewards for the actions and algorithm correctness, 
the maximum number of actions, as well as the maximum number of types $message\_type_{max}$ that an algorithm can have, the resilience factor $r$, and the denominator coefficient $d$ for building the thresholds (cf.\ Section~\ref{section:conditions}). For \textsc{Dave} we define the failure-type to be considered, the system scenarios, and the properties to test the algorithm. This separation of concerns allows \textsc{Dora} to be agnostic to the distributed problem at hand, relying on \textsc{Dave} verification results to learn and improve its performance. \textsc{Dave} is based on SPIN, which has been used to verify many distributed algorithms~\cite{tsuchiya_using_2008, john_towards_2013, an_model_2010}, so our 2-agent system can arguably learn to solve different distributed problems.

After defining the inputs, \textsc{Dora} runs one Learner and $N_G \in \mathbb{N}$ independent Generators. Each Generator executes the generation process to generate a solution for the BGame, i.e., generate a broadcast algorithm. Based on Monte Carlo Tree-Search, our generation process is divided into 4 phases: \emph{Selection}, \emph{Expansion}, \emph{Simulation}, and \emph{Backpropagation}. \LONG{Figure~\ref{fig:monte_carlo_process} shows a visual example of the algorithm. Next, we explain each phase.} 

\subheading{3.2.1 Selection and Expansion}
The generation of the algorithm follows the process defined in Section~\ref{section:playing_the_game}, where \textsc{Dora} chooses actions that change the state of the algorithm to reach the final algorithm. In this first phase, the selection of the action is based on the \emph{Upper Confidence Bound for Trees}~\cite{kocsis_bandit_2006} (UCT) policy. \LONG{This is an improved version of the \emph{UCB}~\cite{sutton_reinforcement_1998} policy applied to tree search problems.} This policy works by choosing the next $action_{selected}$ based on (1) the value and number of visits of $state_{n}$ where $action_{selected}$ leads, and (2) the number of visits in the current $state_{c}$. The equation 
is the following:

\vspace{-0.1cm}
{\tiny
\begin{center}
    $action_{selected}=max(UCT_{action}),\forall action \in actions_{suitable}$ \quad \text{where}
\end{center}
}
{\tiny
\begin{center}
$
UCT_{action} =
\begin{cases}
\dfrac{state_{n}.value}{state_{n}.visits} + C\sqrt{\dfrac{\ln state_{c}.visits}{state_{n}.visits}} & \text{if } state_{n}.visits>0
\\
\infty & \text{if } state_{n}.visits=0
\end{cases}
$
\end{center}
}
\vspace{-0.1cm}

\noindent where $C$ is a constant that manages the algorithm's exploration: a higher value means more exploration. \textsc{Dora} will follow the development process in Section~\ref{section:playing_the_game}, applying this policy to choose the next action, until a \emph{leaf} state is found, i.e., a state never visited before ($state_{n}.visits == 0$). When this happens, the agent moves to the \emph{Expansion} phase. In this phase, the agent will expand the new state found and find its value by running the next phase: the \emph{Simulation} phase.

\LONG{\subsubsection{Expansion}
In this phase, the agent will expand the new state found and find its value by running the next phase: the \emph{Simulation} phase.}

\subheading{3.2.2 Simulation}
\label{sec:simulation}
During the Simulation phase, and starting from the leaf node expanded previously, \textsc{Dora} runs a fixed number of simulations $n_{simulations}$, each generating an algorithm, with the goal of finding the value of the new state discovered. The agent continues to follow the generation algorithm presented in Section~\ref{section:playing_the_game}, where the policy to select the next action is now given by the following expression:


\vspace{-0.1cm}
{\tiny
\begin{center}
$action_{selected} = \begin{cases} random(a), with \ P=0.5 \\ 
min(a.cost), with \ P=0.5 
\end{cases}
\forall a \in actions_{suitable}$
\end{center}
}
\vspace{-0.1cm}
\noindent allowing 
our agents to explore both randomly and by choosing the most efficient actions.

After gathering the information about the expanded state during the \emph{Simulation} phase, \textsc{Dora} executes the \emph{Backpropagation} phase.


\subheading{3.2.3 Backpropagation}
\label{section:learning_process}
After running the \emph{Simulation} phase, the experience obtained is passed to the learning process. In \textsc{Dora}, both the Generator and the Learner components execute the learning process\LONG{(Figure~\ref{fig:solution_architecture})}. Following the MCTS algorithm, this step runs the \emph{Backpropagation} phase\LONG{(Figure~\ref{fig:monte_carlo_process})}. In this phase, each Learner and Generator will update their knowledge by incrementing the $value$ of each state, from the leaf state discovered during the expansion phase to the root node, with the reward of each 
\emph{Simulation}. In each updated state, the $visits$ counter is incremented by $1$ per simulation. If Dave classifies an algorithm as incorrect or unknown, the algorithm is blocked so that it is never generated again. The blocked algorithms are shared between all the Generators.

During the generation process, 
\textsc{Dora} relies on \textsc{Dave} to verify the correctness of the solutions generated, i.e., the algorithms. To achieve that, \textsc{Dave} runs the \emph{Verification Process}.


\heading{3.3 Verification process}
\label{section:verification_process}
\textsc{Dave} is the agent responsible for the \emph{verification}: a process that assesses the correctness of the algorithms generated, i.e., whether it satisfies the properties (Section \ref{section:properties}). \textsc{Dave} contains 
$N_V \in \mathbb{N}$ independent Verifiers, allowing it to verify different algorithms in parallel.

When \textsc{Dave} receives an algorithm, it first accesses the \emph{Database} \LONG{(cf.\ Figure~\ref{fig:solution_architecture})} to check if the algorithm has already been verified. If this check is negative, \textsc{Dave} discharges the
properties that hold \emph{by construction}, avoiding unnecessary model checking. In this work, two
properties are discharged this way: \emph{No-duplication} and
\emph{Integrity2}. \emph{No-duplication} follows from the semantics of
the \instruction{Deliver} instruction (Section~\ref{section:instructions}):
since \instruction{Deliver} terminates the execution of the process, no
process can deliver more than once. \emph{Integrity2} follows from the
failure and network models (Section~\ref{section:system_model}): under
crash and omission failures, processes cannot fabricate values, and the
links are authenticated and integrity-preserving, so any delivered
value is necessarily either the value initiated by the broadcaster or
\variable{default\_value}~$= SF$. Consequently, SPIN only needs to
verify the remaining properties, reducing the verification cost of
each candidate. 
Next, \textsc{Dave} checks the correctness of the algorithm with \emph{SPIN}~\cite{holzmann_model_1997}, a \emph{model checking} framework \cite{clarke_model_2018,baier_principles_2008} often used for verifying fault-tolerant algorithms~\cite{an_model_2010,delzanno_model_2014,john_towards_2013,tsuchiya_using_2008}. 

We use SPIN to simulate the execution of the created algorithm in a specific system model. 
During the state space exploration, SPIN verifies if none of the properties (see Section~\ref{section:properties}) is violated. All the information needed to verify the generated algorithm, such as the algorithm, the values of $N$ and $F$, the system architecture, the behavior of each failure, and the faulty processes (for $F>0$, we always test the broadcaster as faulty, plus one representative assignment among the symmetric non-broadcaster processes) are specified in a PROMELA (Process or Protocol Meta Language) language file (\textit{.pml} extension). In essence, this file is a formal model of a distributed system executing the algorithm under verification. \LONG{Then, based on the \textit{.pml} file, \textsc{Dave} builds a file (\textit{pan.c}), a \textit{C} program that performs a verification of the correctness requirements for the system, and compiles it using \textit{gcc}, generating an executable file.} Then, the verification process uses the SPIN tool to run the executable file, based on the \textit{.pml} model, and, with that, check the correctness of the algorithm. In the end, the result is stored on the Database to be used in future verifications 
without the need for using SPIN, as stated above in this section, and returned to the requesting Generator. 
The verification process considers the \emph{crash} and \emph{general omission} failure models, where we simulate the crash/omission of the process at every possible moment of receiving and/or sending messages following a send order drawn at random for each verified model. When testing, we simulate 
from no process failing to the maximum number of possible failures. The properties of the algorithm are formulated using \emph{Linear Temporal Logic}~\cite{rozier_linear_2011}. 
Note that verifying fault-tolerant distributed algorithms on fixed $(N,F)$ instances via model checking is an established methodology in literature, not a limitation specific to our approach: prior work applies bounded SPIN-based checking to Consensus, Paxos, and other fault-tolerant protocols~\cite{delzanno_model_2014, john_towards_2013,
tsuchiya_using_2008, minamikawa_language_2008}.


\heading{3.4 Reward mechanism}
\label{section:reward_machanism}
\textsc{Dora} learns through \emph{verifiable rewards}, related to the \emph{efficiency} and \emph{correctness} of the solution: most efficient correct solutions will generate better rewards. Rewards are defined as part of the input, but it is important to mention that the agent synthesizes efficient algorithms not due to the absolute reward values that we have defined, but because of the relative values between them 
(the absolute rewards were empirically defined). 


At the end of generating an $algorithm$, \textsc{Dora} receives a $reward$ depending on the correctness of the algorithm:


\vspace{-0.2cm}
{\tiny
\begin{center}
$algorithm_{reward} =
\begin{cases}
algorithm_{penalty} & \text{if } \mathit{isIncorrect}(algorithm) \text{ or } \mathit{isUnknown}(algorithm) \\
-algorithm_{cost} & \text{if } \mathit{isCorrect}(algorithm)
\end{cases}$
\end{center}
}
\vspace{-0.2cm}

The algorithms \textsc{Dora} generates can be classified as \emph{correct}, \emph{incorrect}, or \emph{unknown}. An algorithm is correct if it satisfies all the properties that define the problem (cf. Section~\ref{section:properties}) and is incorrect if it does not satisfy one property. However, \textsc{Dora} may generate algorithms that are too complex for verification in practical time. For that reason, we limit the time that \textsc{Dave} runs (we defined $1$ minute for each property and system configuration verified in the experiments, which allowed the complete verification of more than $95\%$ of the algorithms generated in each test). If time runs out, the algorithm is tagged unknown.

In case (1), where the algorithm is incorrect or unknown, a penalty is given to \textsc{Dora}. Although previous work uses a fixed penalty when the algorithm is wrong~\cite{
vaz_automatic_2023,vaz_synthesis_2023} (e.g., the reward of $-1$ if the algorithm is incorrect), we define a flexible penalty that depends on the number of properties that are not satisfied by the algorithm, which we call \emph{faulty properties}. In essence, the penalty is calculated by:

\vspace{-0.1cm}
{\tiny
\begin{center}
$algorithm_{penalty}$ = $r_{incorrect}\times n_{faulty\_properties} + r_{unknown}\times  n_{unknown\_properties}$
\end{center}
}
\vspace{-0.1cm}

\noindent that multiplies by $r_{incorrect}$ the number of faulty properties ($n_{fault\_properties}$) and by $r_{unknown}$ the number of unknown properties ($n_{unknown\_properties}$). Dead-end candidates are treated as incorrect algorithms violating all properties and receive the maximum penalty. 
%

If the algorithm is correct (case 2), \textsc{Dora} will receive a reward based on the \emph{algorithm's cost}, representing the \emph{efficiency} of the algorithm. More efficient algorithms will have a lower cost, allowing \textsc{Dora} to differentiate between them. In essence, $algorithm_{cost}$ is calculated as:

\vspace{-0.1cm}
{\tiny
\begin{center}
$algorithm_{cost} = \sum action_{cost}, \forall action \in algorithm$
\end{center}
}
\vspace{-0.1cm}

The cost of each action is related to its \emph{efficiency} and is 
influenced by its 3 components: the \emph{instruction}, the \emph{condition}, and the \emph{arguments} used. Next, we analyze all the components.

\subheading{3.4.1 Instructions}
All instructions, except \instruction{Stop}, have a cost, as processes need to spend time and resources to execute them. The \instruction{Deliver} instruction is one of the cheapest instructions, since it allows the algorithm to complete its goal and decide on some value. 
The \instruction{Send to N-1 Neighbors} instruction has a lower cost than \instruction{Send to N Processes} 
due to the number of messages sent: 
\instruction{Send to N Processes} means sending to all, including itself, which represents an increase in computation, even if small, 
compared to \instruction{Send to N-1 Neighbors}. 

\subheading{3.4.2 Conditions}
In terms of conditions, the most efficient condition is \condition{IsTrue} since it is a simple true condition. 
The condition \condition{ReceivedEqualMsg} 
depends on the number of equal messages received and their type. 
Therefore, we define this condition as less efficient.

\subheading{3.4.3 Arguments}
The total cost of an action is also influenced by the \emph{arguments} used by the instruction and the condition. In our work, two arguments can affect the cost of the action: 
the \emph{threshold} and the \emph{message type}. 
The threshold is an argument, only used by conditions, that represents the number of messages needed. The thresholds were ordered, starting with $1$ (the most efficient threshold) and ending with $N$ (the least efficient threshold), using their coefficients. Then, each \variable{threshold} was associated with an increasing cost of $c_{threshold}$, starting from $0$. Formally, $threshold_{cost}=0+c_{threshold}\cdot(order-1)$. For example, if $c_{threshold}=2$, the $1^{st}$ threshold of $1$ gives a cost of $0+2\cdot(1-1)=0$ -- the initial cost -- while the $22^{nd}$ threshold $N$ gives a cost of $0+2\cdot(22-1)=42$. 
In terms of message type, the cost is added both to instructions that can insert a new message type into the algorithm, \instruction{Send} instructions, and to conditions that wait for a specific message type, \condition{ReceivedEqualMsg} conditions. 
For the message type, each type was multiplied by $c_{message\_type}$: for example, if assuming $c_{message\_type}=1$, \type{mtype0} is associated with the cost $0\times 1=0$, \type{mtype1} is associated with the cost $1\times 1=1$, and so on. The objective is for the agent to add the minimum number of new types to the algorithm, as each new type may involve more communication.
\LONG{This multiplication difference between the arguments -- 
$2$ for \variable{threshold}, and $1$ for \variable{message\_type} -- is due to their efficiency importance for the algorithm. 
The \variable{threshold} influences the number of messages that needed to 
make a decision. The \variable{message\_type}, which influences the number of communication steps, was assigned the lowest cost. All reward values were defined following the strategy of previous reward-based works~\cite{vaz_synthesis_2023,vaz_automatic_2023}. Section~\ref{section:sensivity_tests} presents robustness experiments around these values.}

As we have defined, the goal of the agent is to generate a correct algorithm that optimizes multiple criteria: the number of messages sent, the number of messages received, and the number of message types. All of these criteria are influenced by the instructions, conditions, and arguments used. Therefore, 
the goal is not to improve a single criterion, i.e., right/wrong, but to take into account multiple properties and characteristics of the problem. To deal with this, we have implemented a mechanism of \emph{weights}: each component that influences the cost of an action has its weight. In this work, we have defined a  weight for the instruction $w_i$, for the condition $w_c$, and for the argument $w_a$. 
For example, an instruction $A$ that sends more messages must receive a higher cost than an instruction $B$ that sends fewer messages, even if $A$ presents a lower threshold than $B$. Therefore, the cost of an action is: 

\vspace{-0.1cm}
{\scriptsize
\begin{center}
    $action_{cost} = total\_instruction_{cost} + total\_condition_{cost}$ \quad \text{where}
\end{center}
}
{\scriptsize
\begin{center}
$total\_instruction_{cost} =$$instruction_{cost} \times instruction_{weight} + arguments_{cost} \times arguments_{weight}$ \quad \text{and}
\end{center}
}
{\scriptsize
\begin{center}
$total\_condition_{cost} =$$condition_{cost} \times condition_{weight} + arguments_{cost} \times arguments_{weight}$
\end{center}
}

\begin{center}\scriptsize
\text{For example,} \quad if \textcolor{violet}{\condition{ReceivedEqualMsg}}(\variable{v},\textcolor{teal}{\variable{mtype0}},\textcolor{red}{\variable{N}}) do \textcolor{brown}{\instruction{Send to N Processes}}(\variable{v},\textcolor{blue}{\variable{mtype0}},\variable{ID})
\end{center}

{\scriptsize
\begin{center}
\text{would be,} \quad 
$(\textcolor{violet}{(1\times0.10)}$
$+$
$\textcolor{teal}{(0\times1\times0.05)}$
$+$
$\textcolor{red}{(21\times2\times0.05)})$
$+$
$(\textcolor{brown}{(3\times0.85)}+
\textcolor{blue}{(0\times1\times0.05)})=4.75$
\end{center}
}


\heading{3.5 Heuristics}
\label{section:heuristics}

The state space of algorithms grows with the number of selectable actions: more actions mean a bigger space. Without any heuristic, this space would be $\infty$. To limit this explosion, we constrain the search space and guide \textsc{Dora}'s generation process using a set of \emph{heuristics}~\cite{muller_heuristics_1981}. \emph{Note that the heuristics do not help find correct or efficient algorithms; they only reduce the number of states to explore}. At each state, the agent uses heuristics to determine which actions are available, reducing the options from $T$ to $T_h \leq T$. Most heuristics come from related work~\cite{vaz_synthesis_2023}: $9$ of $12$ (H$1$--H$4$, H$7$--H$10$, and H$12$); only H$5$, H$6$, and H$11$ are new. 
H$5$ forbids actions based on variables not allowed in the current event, preventing semantic errors: e.g., in the \event{$\Delta$-Timeout} event, no value \variable{v} is available, so actions depending on it are never executed.
H$6$ enforces an incremental introduction policy on message types: \textsc{Dora} cannot use a higher-indexed type before lower ones have been introduced, e.g., \variable{mtype0} is allowed first, then \variable{mtype0} and \variable{mtype1}, and so on. 
H$11$ forbids any action other than \instruction{Stop} or \instruction{Deliver} after \instruction{Deliver}, since \instruction{Deliver} terminates execution and later actions would likely never run. The full set of heuristics appears in Appendix~\ref{appendix:heuristics}. While heuristics reduce generality, they enable tractable synthesis: our contribution is efficient exploration of the remaining space through MCTS-based learning and flexible reward mechanisms. Section~\ref{section:dora_performance} shows how heuristics reduce the state space without discarding efficient, correct algorithms. In the future, we aim to replace heuristics with a pruning learning process, e.g., discarding partial candidates whose structure resembles previously verified incorrect algorithms.

\vspace{-0.4cm}
\section{Experimental Evaluation}
\label{section:experimental_evaluation}
\vspace{-0.2cm}

Our prototype 
has three software components: 
\textsc{Dora} ($\approx2\,500$ lines of code) implements the learning component using Python $3.10.12$, JSON, and MongoDB; 
\textsc{Dave} ($\approx2\,000$ lines of code) is responsible for the correctness verification and is implemented using Python, JSON, MongoDB, PROMELA, and SPIN; 
the \emph{Library} ($\approx1\,700$ lines of code), implemented in Python and JSON, contains code shared by \textsc{Dora} and \textsc{Dave}.

Our experimental evaluation aimed to address the following questions: 
\smallcircled{Q1} Can \textsc{Dora} and \textsc{Dave} synthesize correct and efficient broadcast algorithms in different failure scenarios? 
\smallcircled{Q2} How do \textsc{Dora} and \textsc{Dave} compare with previous work? To what extent do the use of MCTS, multiple generators, and the reward mechanism contribute to its exploration efficiency? 
\smallcircled{Q3} How does the number of visited states and generated algorithms scale as we increase the number of generators? 
\smallcircled{Q4} How robust is \textsc{Dora} to variations in its parameters?
 
The system was deployed in a High Performance Computing (HPC) system with a total of $144$vCPUs, $288$GB of RAM and $120$GB of disk memory, 
shared between $3$ virtual machines.
All machines were running Ubuntu $22.04$.
The use of $144$ 
vCPUs reflects the inherent cost of formal 
verification ($88\%$ 
of resources were allocated to the verification process). 
This bottleneck is common to model checking verification~\cite{clarke_model_2018, clarke_progress_2001, bellettini_distributed_2013}. Future work will explore learned classifiers to predict correctness, 
and Large Language Model (LLM)-based verification as a complement to model checking.

During the experimental evaluation, we used the following parameter values: $C=54$, empirically defined to match the magnitude of the algorithm rewards (Section 3.4) so that the exploration term is comparable to the exploitation term; $n_{simulations} = 10$, so that the reward of the expanded state is estimated reliably; $r_{incorrect}=-40$, ensuring that no incorrect algorithm can receive a better reward than a correct one; $r_{unknown}=-20$, half of what we assign to incorrect algorithms; $c_{message\_type}=1$ and $c_{threshold}=2$, reflecting their relative importance to the algorithm's efficiency: the \variable{threshold} influences the number of messages needed to make a decision, while the \variable{message\_type} was assigned the lower cost; costs on Table~\ref{table:instructions_conditions} empirically established for each action, inspired by related work~\cite{vaz_synthesis_2023,vaz_automatic_2023}; and $w_{i}=0.85$, $w_{c}=0.10$, and $w_{a}=0.05$, chosen to penalize the active part of an action (the instruction representing behavior) more heavily than its passive part (the condition waiting to be satisfied). Section~\ref{section:sensivity_tests} shows that the results are robust to variations around these values. The Learner shares the global knowledge with the Generators every $120$ algorithms. All the following experiments were capped at $24$ hours per run: a deliberate limit imposed by our available computational budget, rather than a constraint intrinsic to our approach. Even under this budget, \textsc{Dora} and \textsc{Dave} consistently achieved high-quality results. Throughout this section, incorrect algorithms include both candidates rejected by Dave and dead-end candidates.



\begin{table*}
\tiny
\centering 
\caption{Comparison between \textsc{Dora}'s algorithms and human designs, discussing the best case (B) -- no failures -- and worst case (W). We compare in terms of (1) the total number of messages sent, (2) the number of messages required for a process to deliver, and (3) the number of communication steps (i.e., distinct message-sending steps).}
\vspace*{-0.35cm}
\begin{tabular}{ l | c | c | c | c | c | c | c }
\hline 
\textbf{Algorithms} & \textbf{System setting} & \multicolumn{2}{c|}{\textbf{Messages sent}} & \multicolumn{2}{c|}{\textbf{Messages needed}} & \multicolumn{2}{c}{\textbf{Communication Steps}} \\ 
& & B & W & B & W & B & W \\
\hline
\textsc{Dora} Algorithm~\ref{algorithm:best_crash_tolerant_algorithm_generated} & Crash & $O(N^2)$ & $O(N^2)$ & $O(1)$ & \cellcolor{lightgreen!30} $\mathbf{O(1)}$ & $2$ & $N+1$ \\
TRB1
&  failures & $O(N^2)$ & $O(N^3)$ & $O(N)$ & $O(N)$ & $2$ & $N+1$ \\
TRB2
& & $O(N^2)$ & $O(N^2)$ & $O(1)$ & $O(N)$ & $N+1$ & $N+1$ \\
\hline
\textsc{Dora} Algorithm~\ref{algorithm:best_omission_tolerant_algorithm_generated} & Omission & $O(N^2)$ & \cellcolor{lightgreen!30}$\mathbf{O(N^2)}$ & $O(1)$ & \cellcolor{lightgreen!30} $\mathbf{O(1)}$ & $2$ & $N+1$ \\
TRB3
& failures & $O(N^2)$ & $O(N^3)$ & $O(1)$ & $O(N)$ & $2$ & $N+1$ \\
\hline

\end{tabular}
\label{table:algorithm_comparison} 
\end{table*}















\begin{table*} 
\tiny
\centering 
\caption{Summary of the characteristics and statistics of each agent. The difference between each agent and \textsc{Dora} is highlighted in gray.} 
\vspace*{-0.35cm}
\hspace*{-1.2cm}
\begin{tabular}{l | c | 
c | c | c | c | c | c}
\hline 
\textbf{Components}  & 
\textbf{RBLearner} & 
\LONG{\textbf{ROG-Dora} & 
\textbf{SOG-Dora} &}
\textbf{OG-Dora} (\textbf{O}ne & 
\textbf{R-Dora} & 
\LONG{\textbf{RC-Dora} & }
\textbf{C-Dora} & 
\textbf{NEF-Dora} (\textbf{N}ew & 
\textbf{S-Dora} (\textbf{S}tatic  &  
\textbf{Dora} 
\\ 
& \cite{vaz_synthesis_2023,vaz_automatic_2023} & \LONG{(\textbf{R}andom \textbf{O}ne  & 
(\textbf{S}tatic Reward \textbf{O}ne &}
 \textbf{G}enerator \textbf{Dora}) & 
(\textbf{R}andom \textbf{Dora}) & 
\LONG{(\textbf{R}andom- &} 
(\textbf{C}heap \textbf{Dora}) & 
 H\textbf{E}uristic \textbf{F}ree \textbf{Dora})  &   
 Reward \textbf{Dora}) & 
\\
\hline 
Learning Technique &
\cellcolor{gray!30}Q-Learning &
\LONG{\cellcolor{gray!30}Random &
MCTS &}
MCTS &
\cellcolor{gray!30}Random &
\LONG{\cellcolor{gray!30}Random &}
\cellcolor{gray!30}Minimum Cost&
\cellcolor{gray!30}MCTS w/o H5,H6,H11& 
MCTS &
\textbf{MCTS} \\
\hline
Generators &  
\LONG{\cellcolor{gray!30}$1$ &
\cellcolor{gray!30}$1$ &}
\cellcolor{gray!30}$1$ &
\cellcolor{gray!30}$1$ &  
$12$ &
\LONG{$12$ &}
$12$ &
$12$ &
$12$ &
$\mathbf{12}$ \\
\hline
Reward &  
\cellcolor{gray!30}Static &
\LONG{-- &
\cellcolor{gray!30}Static &}
Flexible &
-- &
\LONG{-- &}
-- &
Flexible &
\cellcolor{gray!30}Static &
\textbf{Flexible} \\
\hline 
Unique states visit. &
$615 \pm 109$ &
\LONG{$659 \pm 63$ &
$317 \pm 20$ &}
$286 \pm 21$ &
$4033 \pm 82$ &
\LONG{$1549 \pm 40$ & }
$513 \pm 8$&
$4236 \pm 101$&
$1744 \pm 49$ &
\hspace{-0.15cm}$1657 \pm 91$ \\
\hline
Correct algo. & 
$4 \pm 2$ &
\LONG{\hspace{0.13cm}$5 \pm 1$ &
$12 \pm 3$ &}
$13 \pm 3$ &
\hspace{0.13cm}$15 \pm 2$ & 
\LONG{$45 \pm 5$ & }
\hspace{0.26cm}$0 \pm 0$&
$16 \pm 2$&
\hspace{0.13cm}$65 \pm 7$ &
\hspace{-0.09cm}\cellcolor{lightgreen!30} $\mathbf{72 \pm 7}$\\
\hline
Incorrect algo.  & 
\hspace{-0.13cm}$129 \pm 24$ &
\LONG{$144 \pm 15$ &
$66 \pm 8$ &}
$60 \pm 8$ & 
$1020 \pm 20$ & 
\LONG{$361 \pm 7$ & } 
$498 \pm 7$ &
\hspace{-0.15cm}$1208 \pm 24$ &
\hspace{0.13cm}$391 \pm 19$ &
$356 \pm 20$\\
\hline
Best-3 mean reward &
$-9.06 \pm 2.96$ & 
\LONG{$-7.59$ & 
$-4.53$ &} 
$-4.96 \pm 0.28$ & 
$-6.33 \pm 0.72$ & 
\LONG{$-4.36 \pm 0.11$ & }
-- & 
$-4.55 \pm 0.14$ & 
$-4.36 \pm 0.05$ & 
\hspace{-0.3cm}\cellcolor{lightgreen!30} $\mathbf{-4.32 \pm 0.06}$ 
\\
\hline
Best-5 mean reward &
$-9.47 \pm 2.71$ & 
\LONG{$-8.22$ & 
$-5.04$ &} 
$-5.23 \pm 0.33$ & 
$-6.82 \pm 0.65$ & 
\LONG{$-4.58 \pm 0.12$ & } 
-- &
$-5.25 \pm 0.15$ & 
$-4.60 \pm 0.07$ & 
\hspace{-0.3cm}\cellcolor{lightgreen!30} $\mathbf{-4.46 \pm 0.08}$ 
\\ 
\hline 
\end{tabular}
\label{table:agents_difference} 
\end{table*}

\heading{4.1 Q1 - Generating algorithms}
\label{section:q1_generating_algorithms}

We tested \textsc{Dora} on the synthesis of asynchronous Crash-tolerant Reliable Broadcast (RB) and Uniform Reliable Broadcast (URB), and synchronous Crash- and Omission-tolerant Terminating Reliable Broadcast (TRB). Next, we present the results of each test. 


\subheading{4.1.1 Asynchronous Crash-tolerant RB}
\label{section:async_crash_tolerant_rb}
We follow the system model in Section~\ref{section:system_model}, assuming an asynchronous system with $N \in \mathbb{N}$ processes and $F \in \mathbb{N}$ faulty processes where $F< N$~\cite{cachin_introduction_2011,hadzilacos_modular_1994}. Every algorithm generated was verified in systems $N=3/F=2$ and $N=4/F=3$, since preliminary experiments showed the computational budget to be insufficient for $N\ge5$. We assumed $r=1$ and $d=1$, which allowed generating $8$ thresholds (e.g., $1$, $F + 1$ or $N-F$). Together with a $message\_type_{max}=1$, common in broadcast algorithms~\cite{cachin_introduction_2011,hadzilacos_modular_1994,vaz_synthesis_2023}, resulted in $39$ actions. The problem is defined by properties: \emph{Validity}, \emph{Agreement}, \emph{No-duplication}, \emph{Integrity1}, and \emph{Totality}. We ran \textsc{Dora} for $12$ hours in this scenario. Algorithm~\ref{algorithm:best_crash_tolerant_rb_algorithm} was the most efficient solution generated, equivalent to the version in Hadzilacos and Toueg~\cite{hadzilacos_modular_1994} (Figure 11), where the proof of correctness can be analysed. Throughout this test, \textsc{Dora} generated $128$ correct, $2\,816$ incorrect, and $2$ unknown algorithms, exploring less than $6\%$ ($\approx6\,000$ states) of a space estimated to have more than $10^5$ states (based on mean number of actions \emph{after} heuristics raised to the maximum algorithm size of $8$, as defined in Heuristic 2). Each algorithm took, on average, $0.42\pm(0.08)$ seconds to generate and \textsc{Dora} waited $73.45\pm(95.32)$ seconds for its validation. Two patterns, seen in this test, are common across all the following tests: (1) verification time exhibits large variance, stemming from the behavior of the verification process: some algorithms are quickly rejected, while others require exploring complex simulation paths; and (2) more than $90\%$ of the candidate algorithms evaluated were incorrect, highlighting the intrinsic difficulty, even when using heuristics.


\subheading{4.1.2 Asynchronous Crash-tolerant URB}
We follow the same system model as before: asynchronous system with $N \in \mathbb{N}$ processes and $F \in \mathbb{N}$ faulty processes where $F<N$~\cite{hadzilacos_modular_1994}). 
Every generated algorithm was verified on systems with $N=3/F=2$ and $N=4/F=3$. We assumed the values $r=1$, $d=2$, and $message\_type_{max}=1$, which allowed generating $14$ thresholds (e.g., $1$, $N/2 + 1$ or $F/2 + N/2$), resulting in a total of $63$ actions. The problem was defined by properties: \emph{Validity}, \emph{Uniform Agreement}, \emph{No-duplication}, \emph{Integrity1}, and \emph{Uniform Totality}. We ran \textsc{Dora} for $12$ hours. Throughout this test, \textsc{Dora} generated $31$ correct, $3\,420$ incorrect, and $1$ unknown algorithms, exploring less than $1\%$ ($\approx7\,000$ states) of a space estimated to have more than $10^6$ states. Each algorithm took, on average, $0.65\pm(0.11)$ seconds to generate and \textsc{Dora} waited $62.42\pm(74.97)$ seconds for its validation.

\noindent 
\begin{minipage}[t]{0.48\textwidth}
\vspace{-0.7cm}
\begin{algorithm}[H]
\caption{Best Crash-tolerant RB alg.\ generated by \textsc{Dora}.}
\vspace{-0.1cm}
\label{algorithm:best_crash_tolerant_rb_algorithm}
\begin{tinyalgorithmic}
\State \textbf{// Inner state:}
\State $ID \in {0,...,N-1}$       \Comment{// process identifier}
\State $msg\_received := \varnothing$      \Comment{// set of received messages, initially empty}
\State $msg\_sent := \varnothing$           \Comment{// set of messages sent, initially empty}
\State $delivered\_value := \perp$    \Comment{// initially undefined}
\State $default\_value := SF$
\State 
\State \textbf{// Algorithm:}
\State \event{when Broadcast($v$)}:
\State \hskip0.5em  if \condition{IsTrue()} do \instruction{Send to N Processes($v$,$mtype0$,$ID$)}
\State \hskip0.5em if \condition{IsTrue()} do \instruction{Stop()}
\State \event{end;}
\State \event{when Receive($v$,$t$,$f$) do}: 
\State \hskip0.5em  if \condition{IsTrue()} do \instruction{Send to N-1 Neighbors($v$,$mtype0$,$ID$)}
\State \hskip0.5em  if \condition{ReceivedEqualMsg($v$,$mtype0$,$1$)}  do \instruction{Deliver(\variable{v})}
\State \hskip0.5em if \condition{IsTrue()} do \instruction{Stop()}
\State \event{end;}
\vspace{-0.1cm}
\end{tinyalgorithmic}
\end{algorithm}
\end{minipage}
\hfill
\begin{minipage}[t]{0.48\textwidth}
\vspace{-0.7cm}
\begin{algorithm}[H]
\caption{Best Crash-tolerant URB alg.\ generated by \textsc{Dora}.}
\vspace{-0.1cm}
\label{algorithm:best_crash_tolerant_urb_algorithm}
\begin{tinyalgorithmic}
\State \textbf{// Inner state:}
\State $ID \in {0,...,N-1}$       \Comment{// process identifier}
\State $msg\_received := \varnothing$ \Comment{// set of received messages, initially empty}
\State $msg\_sent := \varnothing$           \Comment{// set of messages sent, initially empty}
\State $delivered\_value := \perp$    \Comment{// initially undefined}
\State $default\_value := SF$
\State 
\State \textbf{// Algorithm:}
\State \event{when Broadcast($v$)}:
\State \hskip0.5em  if \condition{IsTrue()} do \instruction{Send to N Processes($v$,$mtype0$,$ID$)}
\State \hskip0.5em if \condition{IsTrue()} do \instruction{Stop()}
\State \event{end;}
\State \event{when Receive($v$,$t$,$f$) do}: 
\State \hskip0.5em  if \condition{ReceivedEqualMsg($v$,$mtype0$,$1$)} do \instruction{Send to N-1 Neighbors($v$,$mtype0$,$ID$)}
\State \hskip0.5em  if \condition{IsTrue()}  do \instruction{Deliver(\variable{v})}
\State \hskip0.5em if \condition{IsTrue()} do \instruction{Stop()}
\State \event{end;}
\vspace{-0.1cm}
\end{tinyalgorithmic}
\end{algorithm}
\end{minipage}

Algorithm~\ref{algorithm:best_crash_tolerant_urb_algorithm} was the most efficient solution 
generated, equivalent to Algorithm~\ref{algorithm:best_crash_tolerant_rb_algorithm}, 
the crash-tolerant RB algorithm from Section~\ref{section:async_crash_tolerant_rb}. 
Hadzilacos and Toueg~\cite[Theorem~10]{hadzilacos_modular_1994} prove that, under crash 
failures with reliable links (no message loss or corruption, matching our system model, 
Section~\ref{section:system_model}), this algorithm satisfies URB. Intuitively, since every sent completes in full before a process delivers, and links never lose messages, a process cannot deliver without every other process having  already received the same message, meaning the single-message threshold used for RB is already sufficient. This contrasts with the classic \emph{Majority-Ack} URB algorithm of Cachin et al.~\cite{cachin_introduction_2011} (Algorithm~$3.4$): algorithms follow the same flow (the broadcaster disseminates the message, every process relays it upon first reception, and a process delivers after receiving it from a threshold number of processes), but \emph{Majority-Ack} URB requires $\lfloor N/2 \rfloor + 1$ processes, needed to tolerate message loss over unreliable links, whereas Algorithm~\ref{algorithm:best_crash_tolerant_urb_algorithm} 
requires only $1$, achieving optimal resilience for our reliable-link model.


\subheading{4.1.3 Synchronous Crash- and Omission-tolerant TRB} On both failure scenarios, we followed the system model in Section~\ref{section:system_model}, assuming a synchronous system with $N \in \mathbb{N}$ processes and a number of faulty processes $F \in \mathbb{N}$ where $F<N$~\cite{dwork_consensus_1988,raynal_consensus_2002}. The Crash-tolerant case was verified with $N=3/F=2$ and $N=4/F=3$, while the Omission-tolerant case, due to its complexity, was tested with $N=2/F=1$ and $N=3/F=2$. We assumed the values $message\_type_{max}=1$, $r=1$ and $d=3$, which allowed the generation of $22$ thresholds  (e.g., $1$, $N/3 -F/3 + 1$ or $N - 1$), resulting in a total of $95$ actions. The problem was defined by properties: \emph{Validity}, \emph{Agreement}, \emph{No-duplication}, \emph{Integrity2}, \emph{Totality} and \emph{Termination}.

For comparison purposes, we selected the three human-designed TRB algorithms we identified: for the crash-tolerant case, the algorithm of \emph{Best-effort Broadcast} + \emph{Consensus} (Algorithm $6.3$ from~\cite{cachin_introduction_2011} and Figure $16$ from~\cite{hadzilacos_modular_1994}), resulting in the following cases: \emph{Best-effort Broadcast}~\cite{cachin_introduction_2011} + \emph{FloodingSet Consensus}~\cite{cachin_introduction_2011,raynal_consensus_2002,lynch_distributed_1996,attiya_distributed_2004} that we identify here as \emph{TRB1}; and \emph{Best-effort Broadcast}~\cite{cachin_introduction_2011} + \emph{Hierarchical Consensus}~\cite{cachin_introduction_2011} that we identify here as \emph{TRB2}. For the omission case, we consider Roşu's algorithm \cite{rocsu_early_1996}, identified here as \emph{TRB3}. Table~\ref{table:algorithm_comparison} compares human-designed algorithms with those discovered by \textsc{Dora}.

We ran \textsc{Dora} for $24$ hours for each failure scenario. Algorithm~\ref{algorithm:best_crash_tolerant_algorithm_generated} was the most efficient solution generated for the crash-failure setting. Its performance beats those of the TRB1 and TRB2, overtaking them in terms of worst-case message-needed complexity. In both the best and worst cases, any process requires only a single incoming message to deliver the value and terminate. Throughout this test, \textsc{Dora} generated $6$ correct, $4\,817$ incorrect, and $1$ unknown algorithm, exploring less than $1\%$ ($\approx17\,000$ states) of a space estimated to have more than $10^7$ states (based on the actions \emph{after} heuristics). Each algorithm took, on average, $0.99\pm(0.19)$ seconds to generate and \textsc{Dora} waited $105.76\pm(130.83)$ seconds for its validation. 

\noindent
\begin{minipage}[t]{0.48\textwidth}
\vspace{-0.7cm}
\begin{algorithm}[H]
\caption{Best Crash-tolerant TRB alg.\ generated by \textsc{Dora}.}
\vspace{-0.1cm}
\label{algorithm:best_crash_tolerant_algorithm_generated}
\begin{tinyalgorithmic}
\State \textbf{// Inner state:}
\State $ID \in {0,...,N-1}$       \Comment{// process identifier}
\State $msg\_received := \varnothing$      \Comment{// set of received messages, initially empty}
\State $msg\_sent := \varnothing$           \Comment{// set of messages sent, initially empty}
\State $delivered\_value := \perp$    \Comment{// initially undefined}
\State $default\_value := SF$
\State 
\State \textbf{// Algorithm:}
\State \event{when Broadcast($v$)}:
\State \hskip0.5em  if \condition{IsTrue()} do \instruction{Send to N Processes($v$,$mtype0$,$ID$)}
\State \hskip0.5em if \condition{IsTrue()} do \instruction{Stop()}
\State \event{end;}
\State \event{when Receive($v$,$t$,$f$) do}: 
\State \hskip0.5em  if \condition{IsTrue()} do \instruction{Send to N Processes($v$,$mtype0$,$ID$)}
\State \hskip0.5em  if \condition{IsTrue()} do \instruction{Deliver(\variable{v})}
\State \hskip0.5em if \condition{IsTrue()} do \instruction{Stop()}
\State \event{end;}
\State \event{when $\Delta-$Timeout}:
\State \hskip0.5em  if \condition{IsTrue()} do \instruction{Deliver($default\_value$)}
\State \hskip0.5em if \condition{IsTrue()} do \instruction{Stop()} 
\State \event{end;}
\vspace{-0.1cm}
\end{tinyalgorithmic}
\end{algorithm}
\end{minipage}
\hfill
\begin{minipage}[t]{0.48\textwidth}
\vspace{-0.7cm}
\begin{algorithm}[H]
\caption{Best Omission-tolerant TRB alg.\ generated by \textsc{Dora}.}
\vspace{-0.1cm}
\label{algorithm:best_omission_tolerant_algorithm_generated}
\begin{tinyalgorithmic}
\State \textbf{// Inner state:}
\State $ID \in {0,...,N-1}$       \Comment{// process identifier}
\State $msg\_received := \varnothing$      \Comment{// set of received messages, initially empty}
\State $msg\_sent := \varnothing$           \Comment{// set of messages sent, initially empty}
\State $delivered\_value := \perp$    \Comment{// initially undefined}
\State $default\_value := SF$
\State 
\State \textbf{// Algorithm:}
\State \event{when Broadcast($v$)}:
\State \hskip0.5em  if \condition{IsTrue()} do \instruction{Send to N Processes($v$,$mtype0$,$ID$)}
\State \hskip0.5em if \condition{IsTrue()} do \instruction{Stop()}
\State \event{end;}
\State \event{when Receive($v$,$t$,$f$) do}: 
\State \hskip0.5em  if \condition{ReceivedEqualMsg(\variable{v},\variable{mtype0},N-F)} do \instruction{Send to N-1 Neighbors($v$,$mtype0$,$ID$)}
\State \hskip0.5em  if \condition{IsTrue()} do \instruction{Deliver(\variable{v})}
\State \hskip0.5em if \condition{IsTrue()} do \instruction{Stop()}
\State \event{end;}
\State \event{when $\Delta-$Timeout}:
\State \hskip0.5em  if \condition{IsTrue()} do \instruction{Deliver($default\_value$)}
\State \hskip0.5em if \condition{IsTrue()} do \instruction{Stop()}
\State \event{end;}
\vspace{-0.1cm}
\end{tinyalgorithmic}
\end{algorithm}
\end{minipage}

In the omission-failure setting, Algorithm~\ref{algorithm:best_omission_tolerant_algorithm_generated} was the best solution produced by \textsc{Dora}. This algorithm assumes the optimal resilience $F = N-1$, since only then does $N-F=1$, causing the relay threshold on line~$14$ to be satisfied by the same message that triggers \texttt{Deliver} on line~$15$. However, $F$ is only a configuration parameter fixed at synthesis and verification time, defining the worst case the algorithm must tolerate: the algorithm remains correct whenever the number of actual failures occurring during execution is less than or equal to $F$, provided the system itself is configured for $F = N-1$, the optimal resilience. Compared to TRB3, the generated algorithm significantly improves worst-case communication costs, reducing the number of messages sent from $O(N^3)$ to $O(N^2)$, and the number of messages a process must receive from $O(N)$ to $O(1)$. In this experiment, \textsc{Dora} generated $7$ correct and $6\,136$ incorrect algorithms, exploring less than $1\%$ ($\approx21\,000$ states) of a space estimated to have more than $10^7$ states. Each algorithm took, on average, $1.01\pm(0.19)$ seconds to generate and \textsc{Dora} waited $65.21\pm(77.92)$ seconds for its validation. We present the proofs of correctness for both Algorithms~\ref{algorithm:best_crash_tolerant_algorithm_generated} and \ref{algorithm:best_omission_tolerant_algorithm_generated} in Appendix~\ref{appendix:correctness_proofs_2}.
The simplicity of both designed algorithms arises from a flooding-based broadcast strategy, in contrast to the human-designed TRB1/TRB2, which combine consensus primitives. Moreover, this demonstrates \textsc{Dora}'s ability to find efficient solutions that humans may overlook by relying on higher-level building blocks.

\heading{4.2 Q2 - Prior work and effectiveness of components}
\label{section:dora_performance}

Dora's architecture is defined by: 
\circled{1} an MCTS-based learning technique, \circled{2} a \emph{parallel generation} architecture, and \circled{3} a \emph{flexible verifiable reward} mechanism. 
To evaluate the contribution of each component, we implemented and compared $6$ different agents with combinations of these mechanisms.
We compared these agents with 
the \emph{RBLearner} agent~\cite{vaz_synthesis_2023,vaz_automatic_2023} (cf.\ Section \ref{sec:rel-w}), the only previous machine-learning-based solution for the synthesis of distributed algorithms we identified in the literature. RBLearner only synthesizes Reliable Broadcast algorithms, 
so we re-implemented the approach, following the authors' code and configurations, to generate algorithms using identical actions, heuristics, and verification.

Table~\ref{table:agents_difference} summarizes the characteristics of all agents and highlights their differences to \textsc{Dora}. In the experiments, all agents were executed on the same configuration: 
state space, 
heuristics, and 
actions. 
For NEF-Dora, we removed only the heuristics new in this work (H$5$, H$6$, and H$11$), retaining the heuristics we borrowed from \cite{vaz_synthesis_2023,vaz_automatic_2023}. 
We evaluated the ablation using the \emph{no-failure} TRB case ($r=1$, $d=3$, $N=3/F=0$ and $N=4/F=0$), pairing the simplest failure scenario with the most complex state space and synthesis problem. We started with the crash-tolerant RB scenario, but within a 1-hour budget, \textsc{Dora} generated only a mean of $3$ correct algorithms (matching S-\textsc{Dora}, while OG-\textsc{Dora} generated none). A fairer time budget for this scenario would need $4$ hours, where, in a single run, \textsc{Dora} generated $35$ and S-\textsc{Dora} $31$ correct algorithms. Replicating the full ablation design ($5$ variants $\times$ $5$ repeats) at a 4-hour budget for every failure scenario was not feasible given our available compute; we treat the no-failure TRB case, which presents the largest action space and hardest problem among our tested scenarios, as a lower bound on component effects. We aim to extend the ablation study in the future. 
We ran each agent for $1$ hour and repeated $5$ times. For each test, we measured: \emph{number of unique states explored}, \emph{number of correct and incorrect algorithms}, and the \emph{mean reward} of the best $3$ and $5$ algorithms generated.

Table~\ref{table:agents_difference} shows the average results per agent. Comparing \textsc{R-Dora} with \textsc{Dora} ($12$ Generators), MCTS-guided search finds $4.8\times$ more correct algorithms ($15$ vs. $72$) while visiting less than half the states ($4\,033$ vs. $1\,657$), evidencing its sample efficiency. Parallelization (\textsc{OG-Dora} vs. \textsc{Dora}) yields $5.5\times$ more correct algorithms ($13$ vs. $72$), improving the best-3 and best-5 rewards by $13\%$ and $15\%$. The flexible reward mechanism (\textsc{S-Dora} vs. \textsc{Dora}) generates $11\%$ more correct algorithms while exploring $5\%$ fewer states. The new heuristics H5, H6, and H11 (NEF-\textsc{Dora} vs. \textsc{Dora}) reduce the visited states and incorrect algorithms by $61\%$ and $70\%$, while also improving the best-3/best-5 rewards by $5\%$ and $15\%$, indicating they do not eliminate correct and efficient algorithms. Finally, C-\textsc{Dora} shows that heuristics alone do not guarantee correctness: a cost-greedy policy yields only incorrect algorithms.

Compared to related techniques, \textsc{Dora} outperforms the prior machine-learning agent, RBLearner~\cite{vaz_automatic_2023,vaz_synthesis_2023}, generating $3\times$ to $18\times$ more correct algorithms and improving best-3/best-5 mean reward by up to $52\%$/$53\%$, respectively (for $1$--$12$ Generators). Bar-David and Taubenfeld~\cite{bar_automatic_2003} explore more than $10^7$ candidates to solve Mutual Exclusion. Katz and Peled~\cite{katz_genetic_2008,katz_model_2008,katz_synthesizing_2011} evaluate more than $10^5$ programs on average for valid $N=2$ results on Mutual Exclusion and Leader Election. Zieliński~\cite{zielinski_automatic_2007} generates more than $3\times10^7$ candidates but obtains only $6$ crash-tolerant Consensus solutions for $N=5$. Works addressing more general problems tend to restrict the search space. Delporte et al.~\cite{delporte_towards_2019} limit the problem to $N=2$, yielding a small state space ($\approx450$ candidates) and only $2$ correct solutions. Lazić et al.~\cite{lazic_synthesis_2018} synthesize only threshold parameters within a predefined sketch, exploring a reduced design space ($\approx500$ candidates in the hardest case, yielding $4$ correct solutions). In contrast, \textsc{Dora}'s self-learning approach guides exploration, synthesizing approximately $4.5\times10^3$ candidates from a much larger design space and efficiently identifying correct solutions, including some comparable to known algorithms.


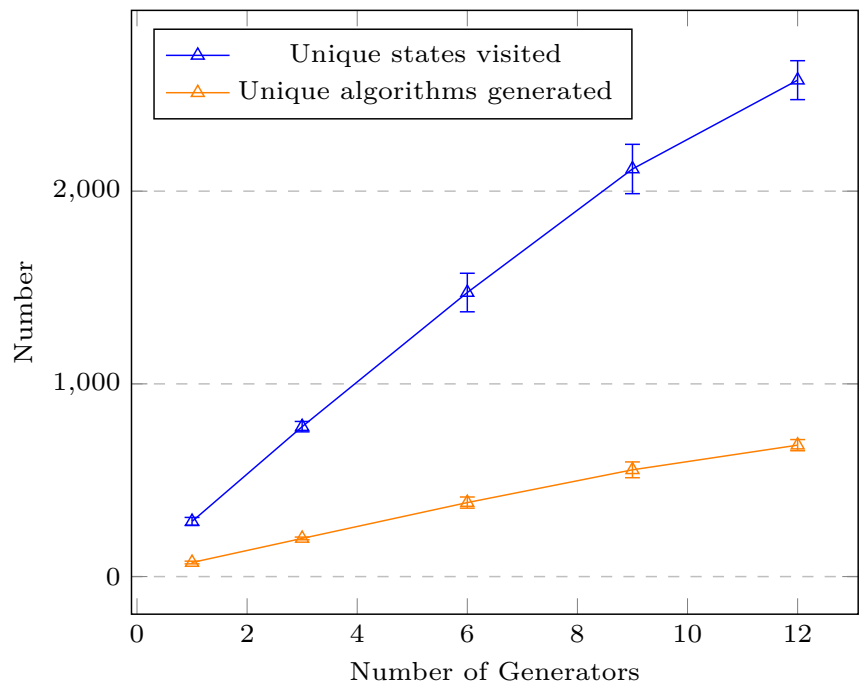
\begin{figure}[th]\scriptsize
\centering
\resizebox{!}{4.5cm}{
\begin{tikzpicture}
\begin{axis}[
    xlabel={Number of Generators},
    ylabel={Number},
    legend pos=north west,
    ymajorgrids=true,
    grid style=dashed,
]


\addplot[
    color=blue,
    mark=triangle,
    error bars/.cd,
    y dir=both,
    y explicit,
]
coordinates {
    (1,286)+-(21, 21)
    (3, 778)+-(27, 27)
    (6, 1474)+-(100, 100)
    (9, 2115)+-(128, 128)
    (12, 2576)+-(101,101)
};


\addplot[
    color=orange,
    mark=triangle,
    error bars/.cd,
    y dir=both,
    y explicit,
]
coordinates {
    (1,73)+-(7, 7)
    (3, 198)+-(7, 7)
    (6, 384)+-(29, 29)
    (9, 554)+-(41, 41)
    (12, 682)+-(29,29)
};


\legend{
Unique states visited, 
Unique algorithms generated,
}

\end{axis}
\end{tikzpicture}
}
\vspace{-0.4cm}
\caption{Unique number of states and algorithms generated based on the number of generators.}
\label{fig:dora_scalability}
\end{figure}

\heading{4.3 Q3 - Scalability analysis}
\label{section:dora_scalability}

We analyzed the state space coverage of \textsc{Dora}, depending on the number of Generators. We have run multiple configurations 
with $1$, $3$, $6$, $9$, and $12$ Generators. Again, we evaluated using the \emph{no-failure} TRB case ($r=1$, $d=3$, $N=3/F=0$ and $N=4/F=0$). Each configuration was run $5$ times. Additionally, across all Dora configurations, we fixed the total number of Verifiers at 12 (each \textsc{Dave} supporting 6 Verifiers), ensuring verification capacity was not a bottleneck as Generators scaled. This explains the higher state coverage for 12 Generators compared with Table 4, where Dora ran with $6$ Verifiers, with each \textsc{Dave} supporting $3$. 


Figure \ref{fig:dora_scalability} presents the average number of unique states visited and unique algorithms generated, showing a consistent increase as the number of Generators grows. Starting from $1$ up to $12$ Generators, the number of unique states visited expands from approximately $286$ to over $2\,500$ states ($\approx9\times$ more states), while the number of unique algorithms generated increases from $73$ to over $680$ ($\approx9.3\times$ more algorithms).
These results show that \textsc{Dora} benefits directly from additional computational resources: increasing the number of Generators yields a near-proportional increase in unique search coverage. As future work, we plan to explore state-partitioning techniques~\cite{segal_scalability_2010,mannion_parallel_2015}, e.g., assigning disjoint regions of the search space to each Generator to increase the benefits of parallelization further.


\begin{table*}
\tiny
\centering
\caption{Summary of the statistics of \textsc{Dora}'s variants evaluating the sensitivity of the parameters.}
\vspace{-0.35cm}
\hspace*{-0.7cm}
\begin{tabular}{l | c | c | c | c | c}
\hline
\textbf{Configuration} & \textbf{Unique states visit.} & \textbf{Correct algo.} & \textbf{Incorrect algo.} & \textbf{Best-3 reward} & \textbf{Best-5 reward} \\
\hline
$C=27$ & $1\,617 \pm 63$ & $58 \pm 10$ & $360 \pm 23$ & $-4.39 \pm 0.08$ & $-4.61 \pm 0.09$ \\
$C=108$ & $1\,544 \pm 37$ & $53 \pm 3$ & $343 \pm 13$ & $-4.39 \pm 0.1$ & $-4.6 \pm 0.13$ \\
\hline
$n=5$ & $1\,396 \pm 51$ & $60 \pm 5$ & $307 \pm 17$ & $-4.32 \pm 0.04$ & $-4.5 \pm 0.06$ \\
$n=20$ & $1\,726 \pm 30$ & $66 \pm 4$ & $374 \pm 9$ & $-4.38 \pm 0.1$ & $-4.58 \pm 0.13$ \\
\hline
$w_{i}=0.70$, $w_{c}=0.20$, $w_{a}=0.10$ & $1\,553 \pm 79$ & $60 \pm 9$ & $339 \pm 28$ & $-3.63 \pm 0.08$ & $-3.85 \pm 0.1$ \\
$w_{i}=0.92$, $w_{c}=0.05$, $w_{a}=0.03$ & $1\,528 \pm 94$ & $59 \pm 4$ & $335 \pm 30$ & $-4.69 \pm 0.05$ & $-4.9 \pm 0.14$ \\
\hline
cost/reward$\times2$ & $1\,630 \pm 49$ & $61 \pm 2$ & $367 \pm 19$ & $-8.75 \pm 0.11$ & $-9.30 \pm 0.1$ 
\\
cost/reward$\times0.5$ & $1\,607 \pm 86$ & $62 \pm 8$ & $353 \pm 16$ & $-2.19 \pm 0.05$ & $-2.30 \pm 0.06$ \\
\hline
\LONG{
$P_{random(a)}=0.25$, $P_{min(a.cost)}=0.75$ & $1\,290 \pm 103$ & $62 \pm 10$ & $296 \pm 19$ & $-4.36 \pm 0.05$ & $-4.56 \pm 0.1$ \\
$P_{random(a)}=0.75$, $P_{min(a.cost)}=0.25$ & $2\,135 \pm 124$ & $52 \pm 4$ & $473 \pm 32$ & $-4.38 \pm 0.08$ & $-4.63 \pm 0.17$ \\
\hline
}
\end{tabular}
\label{table:sensivity_evaluation}
\end{table*}

\heading{4.4 Q4 - Sensitivity evaluation}
\label{section:sensivity_tests}

We analyzed \textsc{Dora}'s parameter sensitivity by scaling their default values -- defined on the beginning of Section~\ref{section:experimental_evaluation} --\LONG{($C=54$, $n=10$, $w_{i}=0.85$, $w_{c}=0.10$, $w_{a}=0.05$, the rewards in Table~\ref{table:instructions_conditions}, $r_{incorrect}=-40$, $r_{unknown}=-20$, and $P_{random(a)}=P_{min(a.cost)}=0.5$)} 
by $2\times$ and $0.5\times$. Evaluation used the \emph{no-failure} TRB case ($message\_type_{max}=1$, $r=1$, $d=3$, $N=3/F=0$ and $N=4/F=0$), with each variant run $5$ times; Table~\ref{table:sensivity_evaluation} shows the results. Varying the UCT exploration constant (from $C=27$ to $C=108$) has limited impact on correctness discovery or solution quality, within variance. Increasing simulations ($n=5$ to $n=20$) improves state-space coverage ($1\,396$ to $1\,726$ unique states) and the number of correct algorithms found ($60$ to $66$), at the cost of more incorrect candidates. Varying the efficiency-cost weights affects rewards more noticeably, reflecting multi-objective trade-offs, while correctness discovery remains comparable (around $60$ correct algorithms). Scaling the instruction costs and penalties by $2\times$ and $0.5\times$ scales rewards roughly proportionally, as expected, but leaves the number of correct algorithms essentially unchanged, indicating that \textsc{Dora} does not require brittle reward tuning; future work could normalize rewards to a smaller range, e.g., $[0,1]$, following common practice~\cite{kocsis_bandit_2006,browne_survey_2012}. \LONG{Finally, simulation-phase probabilities affect exploration: biasing toward random choices ($P_{random(a)}=0.75$) visits $66\%$ ($2\,135$ vs. $1\,290$) more unique states than biasing toward cost-effective choices ($P_{random(a)}=0.25$), yet the number of correct algorithms and their rewards are essentially unchanged.} Overall, \textsc{Dora}'s output is robust to $2\times$/$0.5\times$ perturbations: only the exploration/exploitation balance is affected, without degrading quality.

\vspace{-0.4cm}
\section{Related Work}
\label{sec:rel-w}
\vspace{-0.2cm}

Automatic program generation has long been studied as \emph{Program Synthesis}~\cite{gulwani_program_2017}, applied to domains ranging from security protocol generation~\cite{song_agvi_2001} to network control-plane operations~\cite{campbell_avenir_2021}. Within distributed computing, prior works targeted single problems and systems: 
failure-free asynchronous Mutual Exclusion~\cite{bar_automatic_2003,katz_genetic_2008,katz_model_2008}, crash-stop and Byzantine-tolerant asynchronous Consensus~\cite{zielinski_automatic_2007}, failure-free asynchronous Consensus~\cite{delporte_towards_2019}, failure-free asynchronous Leader Election~\cite{katz_synthesizing_2011}, or asynchronous Byzantine-tolerant Reliable Broadcast and one-step Consensus~\cite{lazic_synthesis_2018}. Closest to ours, a reward-based Reinforcement Learning agent~\cite{vaz_synthesis_2023,vaz_automatic_2023} synthesizes only Reliable Broadcast on an asynchronous system. In contrast, our system supports: \circled{1} multiple Broadcast variants, based on a set of properties; \circled{2} two different failure scenarios (crash or general-omission); and \circled{3} both synchronous and asynchronous communication.

Beyond the problems above, prior work explored brute force~\cite{bar_automatic_2003}, model-checking-based genetic programming~\cite{katz_model_2008,katz_genetic_2008,katz_synthesizing_2011}, enumerative~\cite{zielinski_automatic_2007}, SMT-based model checking~\cite{delporte_towards_2019}, and CEGIS~\cite{lazic_synthesis_2018} techniques. However, the combinatorial explosion of candidate algorithms and execution states~\cite{bar_automatic_2003,zielinski_automatic_2007,delporte_towards_2019} forces these techniques to restrict the search space by assuming partially complete algorithms~\cite{lazic_synthesis_2018}, optimistic assumptions~\cite{zielinski_automatic_2007}, pre-defined correct algorithms~\cite{katz_genetic_2008}, or non-parametric designs tailored to a fixed system size~\cite{bar_automatic_2003,katz_model_2008,delporte_towards_2019}.
Relative to the prior machine learning system~\cite{vaz_automatic_2023,vaz_synthesis_2023}, 
our approach introduces three innovations: \circled{1} an MCTS-based learning technique, \circled{2} a parallel generation architecture, and \circled{3} a flexible verifiable reward mechanism. We compare against it in Section~\ref{section:experimental_evaluation} and significantly improve over that work.

Outside distributed algorithms, recent work on neural and LLM-guided program synthesis increasingly combines generative models with search and external verification. VerMCTS integrates an LLM with MCTS, using a logical verifier to evaluate partial programs and guide search toward verified solutions \cite{brandfonbrener2024vermcts}. Similar ideas appear in formal theorem proving: DeepSeek-Prover-V1.5 combines reinforcement learning from proof-assistant feedback with MCTS-based search \cite{xin2025deepseekprover}, while AlphaProof uses a learned policy and value function to guide tree search in the formally verifiable Lean environment \cite{hubert2026alphaproof}. These approaches illustrate a broader trend toward neuro-symbolic synthesis, where learned models guide search while formal tools provide correctness feedback. In contrast, \textsc{Dora} does not rely on an LLM or a pretrained neural model to generate candidate algorithms: it performs MCTS over a structured space of distributed algorithms, while \textsc{Dave} verifies the resulting candidates. Beyond the training bias and computational cost discussed, this design also avoids the unpredictability of unrestricted neural generation, while retaining the benefits of formally-guided search.

Ensuring correctness requires formal validation: \emph{Program Verification}~\cite{fetzer_program_1988}, and model checking in particular, have a long history analyzing fault-tolerant distributed algorithms~\cite{an_model_2010,
delzanno_model_2014,goel_towards_2021,john_towards_2013,
lamport_specifying_1994,minamikawa_language_2008,tsuchiya_using_2008,rahli_velisarios_2018}, with tools including SPIN~\cite{holzmann_model_1997}, ByMC~\cite{konnov_bymc_2018}, and TLA+~\cite{lamport_specifying_2002}. ByMC's parametric verification is limited to threshold-based algorithms, restricting future guard conditions, e.g., role-based conditions such as \condition{if I am the coordinator}~\cite{vaz_2025_ai,schiper_early_1997}; we therefore adopted SPIN, a mature and flexible framework successfully used to verify Consensus~\cite{tsuchiya_using_2008}, Reliable Broadcast~\cite{john_towards_2013}, and Leader Election~\cite{an_model_2010}.

\vspace{-0.4cm}
\section{Conclusion and Future Work}
\vspace{-0.2cm}

We presented a new test-time verifier-guided 
approach to generate distributed algorithms, based on two agents: \textsc{Dora} and \textsc{Dave}. Our approach synthesizes correct broadcast algorithms competitive with human designs for RB, URB, and TRB, demonstrating generality across problems. Our system currently targets crash and general-omission failures under
a fixed three-event template, and verifies small, fixed $(N,F)$
instances. We plan to: 
\circled{1} extend \textsc{Dave} to more failure and timing models (e.g., Byzantine failures, partial synchrony) and problems (e.g., Consensus, Leader Election);
\circled{2} make \textsc{Dora}'s template and action set more flexible;
\circled{3} optimize verification, e.g., classification solutions;
\circled{4} generalize verification to the parametric case, by proving correctness for all $N$ from a bounded set of verified instances, following Konnov et al.'s work~\cite{konnov_short_2017}; and
\circled{5} replace heuristics with a learned pruning process that discards candidates structurally similar to previous generated incorrect algorithms.

\bibliography{bibliography}

 
\appendix

\section{Building an Algorithm}
\label{appendix:building_an_algorithm}

\begin{algorithm}
\caption{Pseudo-code for playing the game, i.e., generating an algorithm.}
\label{algorithm:playing_the_bcgame}
\begin{algorithmic}[1]
\tiny
\Function{GenerateAlgorithm}{}
\State $algorithm \gets [[],[],[]]$
\State $i_{event} \gets 0$
\While{$True$}
    \State $actions_{all} \gets getAllActions()$
    \State $actions_{suitable} \gets applyHeuristics(actions_{all}, algorithm)$
    \State $actions_{suitable} \gets removeBlocked(actions_{suitable}, algorithm)$ \Comment{discard blocked algorithms}
    \If{$actions_{suitable} = \emptyset$}
      \State \Return $algorithm$ \Comment{dead end}
    \EndIf
    \State $action_{selected} \gets getNextAction(actions_{suitable})$
    \State $algorithm[i_{event}] \gets algorithm[i_{event}] + action_{selected}$ \Comment{add action to algorithm}
    \If{$isComplete(algorithm)$}
      \State \Return $algorithm$ \Comment{generation is over}
    \ElsIf{$isStop(action_{selected})$}
      \State $i_{event} \gets i_{event} + 1$ \Comment{move to next event}
    \EndIf
\EndWhile
\EndFunction
\end{algorithmic}
\end{algorithm}

Dora generates an algorithm by adding actions to the template until a final solution is reached. The pseudo-code for generating an algorithm is represented by Algorithm~\ref{algorithm:playing_the_bcgame}. The game starts with the algorithm (variable $algorithm$) empty (line $2$), as defined in Algorithm~\ref{algorithm:algorithm_structure}. Dora will start developing the first event of the algorithm (line $3$): the Broadcast event. Initially, the agent collects $actions_{all}$: a list with all possible actions (line $5$). Then, based on the current state of $algorithm$ and the list $actions_{all}$, Dora analyzes a set of heuristics (line $6$), a topic we debate in Section~\ref{section:heuristics}. From that analysis, the heuristics will discard actions that are not suitable for the current state, resulting in a list $actions_{suitable} \subseteq actions_{all}$. Dora then removes the actions that would lead to a blocked algorithm, i.e., one previously classified by Dave as incorrect or unknown (line $7$). If no action remains, the generation ends in a dead end (lines $8$--$9$). From that new list $actions_{suitable}$, the agent selects an $action_{selected}$ (line $10$), adding it to the current algorithm (line $11$). Then, Dora checks if the algorithm is complete (line $12$). Completeness is achieved when all events contain the Stop action. As presented above, our board/algorithm is composed of three events: Broadcast, Receive, and $\Delta$-Timeout. The generation of the algorithm starts in the Broadcast event, where actions are added to the event until a Stop action is chosen (line \new{$14$}). When this happens, the agent moves to the following event (line \new{$15$}), the Receive event, and continues to add actions until, again, the Stop action is chosen. Then, the agent will continue to choose actions, but now added to the last event of the algorithm, the $\Delta$-Timeout event, added until Stop is chosen again. When all 3 events have the Stop action, the algorithm is considered finished. If that happens, the algorithm is complete and the generation process stops (line \new{$13$}). Otherwise, the agent continues generating the algorithm (returns to line $5$) and adds actions to the current event under development.


\section{Heuristics}
\label{appendix:heuristics}

Dora uses a set of heuristics to prune the state space, allowing it to prevent the exploration of incorrect algorithms. The total set of heuristics is presented in Table~\ref{tab:heuristics}.

\begin{table}\scriptsize
  \caption{All heuristics used to guide the generation process.}
  \label{tab:heuristics}
  \hspace*{-1.2cm}
  \centering
  \begin{tabular}{l | l}
  \hline
    \textbf{Heuristic} & \textbf{Description} \\
    \hline
    H1 &
      Says that events cannot contain duplicate actions on the same event. \\
    H2 &
      Allows for restricting the maximum size of the algorithm generated in terms of \\
      & the number of actions, defined in this work as $8$: the same value used in previous work~\cite{vaz_synthesis_2023}. \\
    H3 &
      Requires algorithms to have at least one \instruction{Deliver} and one \instruction{Send} action, as these actions must clearly  \\
      & exist in the algorithm: without \instruction{Deliver}, no process ever delivers and terminates; without \instruction{Send}, \\
      & no message is ever shared.  \\
    H4 &
      Forbids the agent to select actions based on invalid conditions, e.g., we forbid the agent to select \\
      & actions that wait for message types not yet contained in the algorithm.\\
    H5 &
      Forbids selecting actions based on variables not allowed in the current event, preventing semantic errors.  \\
      & For example, in the \event{$\Delta$-Timeout} event, there is no value \variable{v} available, which means that actions \\
      & depending on \variable{v} are never executed. \\
    H6 &
      Restricts the use of message types to enforce an incremental introduction policy. Specifically, \textsc{Dora} is \\
      & prevented from using higher-indexed message types before lower ones have been introduced. \\
      & For example, if no message type is sent, we only allow sending the message type \variable{mtype0}. \\
      & Then, we only allow sending the \variable{mtype0} and \variable{mtype1}, and so on.  \\
    H7 &
      Does not allow the algorithm to have actions on the \event{Broadcast} event that wait for messages. \\ 
      & Such actions will never be executed, since the algorithm only receives messages on the \event{Receive} event. \\
    H8 &
      Restricts the sending of messages within the same event for each condition and arguments that logically \\ 
      &  overlap. For example, if the current event already contains the action \condition{if IsTrue()} \\ 
      & \instruction{do Send to N Processes}(\variable{v},\variable{mtype0},\variable{ID}), it makes no sense to add the action \condition{if IsTrue()} \\ 
      & \instruction{do Send to N-1 Neighbors}(\variable{v},\variable{mtype0},\variable{ID}), since the first action already sends messages to N, \\
      & with the same message type, upon the same condition.  \\
    H9 &
      Does not allow \textsc{Dora} to generate an algorithm with empty events, as we believe that each event is important. \\
    H10 &
      Forces \textsc{Dora} to choose only actions that send the value \variable{v} and the message type \variable{mtype0} in the event \\ 
      & \event{Broadcast}, since this is the event that starts the algorithm. \\
    H11 &
      Forbids the agent from selecting an action different from \instruction{Stop} or \instruction{Deliver} after selecting an action \instruction{Deliver}. \\ 
      & Since the action \instruction{Deliver} terminates the execution of the algorithm, there is a high probability \\
      & that the actions next to \instruction{Deliver} will never be executed. \\
    H12 &
      Forces the \event{Broadcast} event to contain a \instruction{Send} instruction, since this is the starting event, and \instruction{Send} \\
      & is the only way to share the initial value. \\
      \hline
  \end{tabular}
\end{table}


\section{Correctness Proofs for Algorithms~\ref{algorithm:best_crash_tolerant_algorithm_generated} and \ref{algorithm:best_omission_tolerant_algorithm_generated}}
\label{appendix:correctness_proofs_2}
 
We prove that Algorithm~\ref{algorithm:best_crash_tolerant_algorithm_generated} (synchronous Crash-tolerant TRB) and Algorithm~\ref{algorithm:best_omission_tolerant_algorithm_generated} (synchronous Omission-tolerant TRB) satisfy the six properties that define TRB (Section~\ref{section:properties}): \emph{Validity}, \emph{Agreement}, \emph{No-duplication}, \emph{Integrity2}, \emph{Totality}, and \emph{Termination}. Both algorithms work under the system model of Section~\ref{section:system_model}: $N$ processes, fully connected, synchronous, with reliable, authenticated, integrity-preserving links (no loss, duplication, corruption, or impersonation; reordering allowed). Exactly one process invokes \event{Broadcast($v_0$)} per execution, and up to $F < N$ processes are faulty, consistent with every configuration evaluated in Section~\ref{section:q1_generating_algorithms} ($N{=}3/F{=}2$, $N{=}4/F{=}3$), i.e., at least one process is guaranteed correct, for the crash-tolerant case and $N{=}2/F{=}1$, $N{=}3/F{=}2$ for the omission-tolerant scenario. A process is \emph{correct} if it never crashes, on Algorithm~\ref{algorithm:best_crash_tolerant_algorithm_generated}, and if it suffers no send- and no receive-omission, on Algorithm~\ref{algorithm:best_omission_tolerant_algorithm_generated}. The proofs are organized by property for both Algorithms~\ref{algorithm:best_crash_tolerant_algorithm_generated} and ~\ref{algorithm:best_omission_tolerant_algorithm_generated}.

\begin{remark}[Scope of Algorithm~\ref{algorithm:best_omission_tolerant_algorithm_generated}]
Correctness of Algorithm~\ref{algorithm:best_omission_tolerant_algorithm_generated} below is established for $F = N-1$, the optimal resilience under the general-omission model. Under this constraint, a process's first received message simultaneously satisfies the relay condition $\mathit{ReceivedEqualMsg}(v,\mathit{mtype0},N-F)$ on line~14 (since $N-F=1$) and the unconditional \textsc{Deliver} on line~15, so the two actions trigger together. For $F < N-1$, \textsc{Deliver} would fire on the first received message before the relay threshold $N-F>1$ could be reached, making the relay unreachable; the algorithm would then fail Totality in general, since a correct process's relay would never occur.

\label{remark_1}
\end{remark}

\subsection{Preliminary Lemmas}
 
\begin{lemma}[Atomic termination]
\label{lemma:atomic_termination}
Events execute atomically and in program order. Consequently, in the \event{Receive} event of both Algorithms~\ref{algorithm:best_crash_tolerant_algorithm_generated} and~\ref{algorithm:best_omission_tolerant_algorithm_generated}, the \instruction{Send} action always completes before \instruction{Deliver} executes, and once \instruction{Deliver} executes, the algorithm terminates for that process: no later message and no pending \event{$\Delta$-Timeout} triggers any further instruction.
\end{lemma}
 
\begin{proof}
This is definitional from the model of Sections~\ref{section:events}--\ref{section:instructions}: ``while an event is being executed, no other event can be processed,'' and the \instruction{Deliver} instruction ``allows the process to deliver a message $\ldots$ and terminates the execution of the algorithm.'' Since the \instruction{Send} action is listed strictly before \instruction{Deliver} in both Algorithms~\ref{algorithm:best_crash_tolerant_algorithm_generated} and~\ref{algorithm:best_omission_tolerant_algorithm_generated} \event{Receive} handler, the send always completes first.
\end{proof}
 
\begin{corollary}
\label{cor:one_deliver}
Every process calls \instruction{Deliver} at most once per execution.
\end{corollary}
 
\begin{lemma}[Single-origin value]
\label{lemma:single_origin}
In an execution with exactly one \event{Broadcast($v_0$)} call, every \type{mtype0} message ever sent by any process (broadcaster or not) carries the value $v_0$. Hence every value delivered on the \event{Receive} event, on both Algorithms~\ref{algorithm:best_crash_tolerant_algorithm_generated} and~\ref{algorithm:best_omission_tolerant_algorithm_generated}, by the \instruction{Deliver} instruction equals $v_0$.
\end{lemma}
 
\begin{proof}
 
The first \type{mtype0} message of the algorithm execution is the broadcaster's \instruction{Send to N Processes}($v_0$,\type{mtype0},$ID$) inside \event{Broadcast($v_0$)}. 
Any other relay is executed inside some \event{Receive($v$,$t$,$f$)} event (the only other place an \type{mtype0} send occurs), forwarding the value $v$ of the message that triggered it. Because links are authenticated and integrity-preserving (no impersonation, fabrication, or corruption), that triggering message must be a genuine copy of an earlier \type{mtype0} send, whose value is $v_0$, equal to the first \type{mtype0} message sent by the  broadcaster during the \event{Broadcast} event. So $v=v_0$, triggering \instruction{Deliver}($v$) inside \event{Receive}.
\end{proof}
 
\subsection{Validity}
\emph{If a correct process broadcasts $v$, then (1) it eventually delivers some value and (2) the value does not differ from $v$.}

\begin{proof}
(Algorithm~\ref{algorithm:best_crash_tolerant_algorithm_generated}) Let the broadcaster $p$ be correct (never crashes). $p$ executes \event{Broadcast($v_0$)} to completion, so \instruction{Send to N Processes}($v_0$,\type{mtype0},$ID_p$) reaches all $N$ destinations, including $p$ itself (links are reliable and $p$ does not crash mid-send). $p$ therefore later executes \event{Receive($v_0$,$mtype0$,$ID_p$)} and, by Lemma~\ref{lemma:atomic_termination} and \ref{lemma:single_origin}, executes \instruction{Deliver}($v_0$): it eventually delivers (part~1) a value equal to $v_0$ (part~2).
\end{proof}

 \begin{proof}
(Algorithm~\ref{algorithm:best_omission_tolerant_algorithm_generated}) Identical argument, with ``correct'' meaning no send/receive omission: $p$'s multicast to all $N$ processes (including itself) suffers no omission, and $p$ suffers no receive-omission either, so $p$ receives its own message and delivers $v_0$. 
\end{proof}

\subsection{Integrity2}
\label{section:integrity2_proof}
\emph{If a correct process delivers $v$, then $v=SF$ or $v$ was broadcast by some process.}

\begin{proof}
Immediate from Lemma~\ref{lemma:single_origin}: any value delivered via \event{Receive} equals $v_0$, the value actually broadcast by the unique broadcaster; any value delivered via \event{$\Delta$-Timeout} equals the constant $default\_value = SF$. So every delivered value is either $SF$ or the broadcast value, for both Algorithms~\ref{algorithm:best_crash_tolerant_algorithm_generated} and~\ref{algorithm:best_omission_tolerant_algorithm_generated}.
\end{proof}

\subsection{Agreement}
\emph{If two correct processes deliver, they do not deliver distinct values.}

\begin{claim}[Algorithms ~\ref{algorithm:best_crash_tolerant_algorithm_generated} and~\ref{algorithm:best_omission_tolerant_algorithm_generated}]
If correct processes $p$, $q$ deliver $v'$, $v''$, then $v'=v''$.
\end{claim}
 
\begin{proof}
By an immediate corollary of \emph{Integrity2} (Section~\ref{section:integrity2_proof}), $v',v'' \in \{v_0,SF\}$ ($v_0$ if delivered via \event{Receive}, $SF$ if delivered via \event{$\Delta$-Timeout}). Suppose, for contradiction, $v' \ne v''$; $p$ delivered $v_0$ via \event{Receive} and $q$ delivered $SF$ via \event{$\Delta$-Timeout} (meaning $q$'s local timer expired before it received anything relevant).
 
Since $p$ delivered $v_0$ via \event{Receive}, by Lemma~\ref{lemma:atomic_termination} it first executed \instruction{Send to N Processes} \instruction{($v_0$,\type{mtype0},$ID_p$)} in Algorithm~\ref{algorithm:best_crash_tolerant_algorithm_generated}, or, under the $F=N-1$ assumption of Remark~\ref{remark_1},
\instruction{Send to N-1 Neighbors}($v_0$,\type{mtype0},$ID_p$) in Algorithm~\ref{algorithm:best_omission_tolerant_algorithm_generated} \emph{before} delivering. Because $p$ is correct, this send is never interrupted or partially omitted, so it reaches every other process, in particular $q$.
 
Because $q$ is also correct, it is guaranteed to receive this message (no crash or receive-omission before processing it). 

Recall (Section \ref{section:events}) that every process arms its $\Delta$\event{-Timeout} at the common start time $T_0$ and that $\Delta$ covers the algorithm's full worst-case execution, so $q$ receives $p$'s relay before $q$'s own \event{$\Delta$-Timeout} can fire. Hence $q$'s \event{Receive} event triggers first and, by Corollary~\ref{cor:one_deliver}, $q$ delivers via \event{Receive} (value $v_0$, by Lemma~\ref{lemma:single_origin}), contradicting $q$ delivering $SF$ via timeout.
 
Therefore $v'=v''$: Agreement holds for both Algorithms~\ref{algorithm:best_crash_tolerant_algorithm_generated} and~\ref{algorithm:best_omission_tolerant_algorithm_generated}.
\end{proof}

\subsection{No-duplication}
\emph{If a correct process delivers $v$, it does not deliver $v$ more than once.}

\begin{proof}
Immediate from Corollary~\ref{cor:one_deliver}: every process, correct or not, calls \instruction{Deliver} at most once per execution. Holds identically for both Algorithms~\ref{algorithm:best_crash_tolerant_algorithm_generated} and~\ref{algorithm:best_omission_tolerant_algorithm_generated}, independent of the failure model, since it follows purely from Lemma~\ref{lemma:atomic_termination}.
\end{proof}

\subsection{Termination}
\label{appendix:termination}
\emph{Every correct process eventually delivers some value.}

\begin{proof}
Fix an arbitrary correct process $q$ (either Algorithm~\ref{algorithm:best_crash_tolerant_algorithm_generated} or ~\ref{algorithm:best_omission_tolerant_algorithm_generated}). $q$'s local \event{$\Delta$-Timeout} is a deterministic local clock event, unaffected by message loss or omission, and (being correct) $q$ executes every event assigned to it to completion. Two exhaustive cases:
\begin{itemize}
  \item If $q$ receives some message before its local $\Delta$ elapses, its \event{Receive($v$,$t$,$f$)} event fires; both algorithms' \event{Receive} handler contains \condition{IsTrue()} $\to$ \instruction{Deliver}($v$) 
  , which $q$ executes fully (it is correct), so $q$ delivers.
  \item Otherwise, $q$'s \event{$\Delta$-Timeout} fires at time $\Delta$ and $q$ executes \instruction{Deliver}($default\_value$) $=$ \instruction{Deliver}($SF$).
\end{itemize}
Either way $q$ delivers exactly once (Corollary~\ref{cor:one_deliver}), within bounded time $\Delta$. Since $q$ was arbitrary, every correct process eventually delivers, and Termination holds for Algorithm~\ref{algorithm:best_crash_tolerant_algorithm_generated} and ~\ref{algorithm:best_omission_tolerant_algorithm_generated}.
\end{proof}

\subsection{Totality}
\emph{If one correct process delivers a value, then all correct processes eventually deliver some value.}

\begin{proof}
This follows as an immediate corollary of \emph{Termination} (Section~\ref{appendix:termination}), which shows every correct process delivers unconditionally, regardless of whether any other process has delivered, for both Algorithms ~\ref{algorithm:best_crash_tolerant_algorithm_generated} and~\ref{algorithm:best_omission_tolerant_algorithm_generated}.
\end{proof}

\end{document}